# Energy Efficient UAV-Based Service Offloading over Cloud-Fog Architectures

**HATEM A. ALHARBI[1], BARZAN A. YOSUF[2], MOHAMMAD ALDOSSARY[3], JABER ALMUTAIRI[4], JAAFAR M. H. ELMIRGHANI[2], (Fellow, IEEE)**

[1]Department of Computer Engineering, College of Computer Science and Engineering, Taibah University, Madinah 42353, Saudi Arabia.
[2]School of Electronic and Electrical Engineering, University of Leeds, Leeds, LS2 9JT, U.K.
[3]Department of Computer Science, College of Arts and Science, Prince Sattam bin Abdulaziz University, Al-Kharj 16278, Saudi Arabia.
[4]Department of Computer Science, College of Computer Science and Engineering, Taibah University, Madinah 42353, Saudi Arabia.

Corresponding author: Hatem A. Alharbi (hmehmadi@taibahu.edu.sa).

**ABSTRACT** Unmanned Aerial Vehicles (UAVs) are poised to play a central role in revolutionizing future services offered by the envisioned smart cities, thanks to their agility, flexibility, and cost-efficiency. UAVs are being widely deployed in different verticals including surveillance, search and rescue missions, delivery of items, and as an infrastructure for aerial communications in future wireless networks. UAVs can be used to survey target locations, collect raw data from the ground (i.e., video streams), generate computing task(s) and offload it to the available servers for processing. In this work, we formulate a multi-objective optimization framework for both the network resource allocation and the UAV trajectory planning problem using Mixed Integer Linear Programming (MILP) optimization model. In consideration of the different stake holders that may exist in a Cloud-Fog environment, we minimize the sum of a weighted objective function, which allows network operators to tune the weights to emphasize/de-emphasize different cost functions such as the end-to-end network power consumption (EENPC), processing power consumption (PPC), UAV's total flight distance (UAVTFD), and UAV's total power consumption (UAVTPC). Our optimization models and results enable the optimum offloading decisions to be made under different constraints relating to EENPC, PPC, UAVTFD and UAVTPC which we explore in detail. For example, when the UAV's propulsion efficiency (UPE) is at its worst (10% considered), offloading via the macro base station is the best choice and a maximum power saving of 34% can be achieved. Extensive studies on the UAV's coverage path planning (CPP) and computation offloading have been conducted, but none has tackled the issue in a practical Cloud-Fog architecture in which all the elements of the access, metro and core layers are considered when evaluating the service offloading in a distributed architecture like the Cloud-Fog.

## I. INTRODUCTION

The deployment of smart cities is gaining momentum as it is currently being pursued by major private and public sectors across the world. To deliver on the envisioned promises of smart city, there is a vital need to embrace a wide range of Internet connected devices that include wearables, appliances, connected vehicles, CCTV cameras, embedded sensors etc., which are expected to be deployed in very large numbers in different applications such as industrial, health, logistics, energy, etc. This leads us to the realm of Internet of Everything (IoE) [1]. As a result, the number of Internet connected devices is projected to reach 500 billion by 2025, generating sheer amounts of data at unprecedented rates [2].

Among the Internet connected devices, Unmanned Aerial Vehicles (UAVs), widely known as drones are emerging as indispensable tools for a myriad of applications including but not limited to surveillance, delivery of items, search and rescue missions, ad-hoc communication systems for battery constrained devices in future wireless networks, etc. [3]. The driving factors behind the popularity of drones in recent years stem from the maturity of the technology in terms of their light weight, miniature size, and cost-efficiency [4], [5]. UAVs are equipped with different sensors and communication interfaces that may be used to collect and communicate data about the physical world [6]. For instance, high-definition video cameras can be used to stream videos of different points of interests during a flight trip to a ground controller for intelligence purposes. Traditionally, in most verticals, the ground controller (or the base station/gateway) forwards the collected data over multiple networking hops towards a cloud data center for centralized processing [7]. However, transporting all kinds of tasks to the cloud has its own drawbacks in terms of increased power consumption due to the number of nodes activated between the end devices and the cloud data center, added latency due to the long distance and the already congested core network. Hence, this impacts decision making which is unsuitable for mission critical applications and privacy due to the offloading of sensitive data to different cloud vendors [8].

On the other hand, fog computing is a decentralized paradigm that aims at tackling the limitations introduced by the centralized cloud approach. With fog, cloud services such as Infrastructure as a Service (IaaS), Platform as a Service (PaaS), and Software as a Service (SaaS) can be extended from the core network all the way to the edge network where the end-devices are deployed [9]. More precisely, both edge and core networking equipment sites such as sites that host end-devices, base stations, Wi-Fi, routers, and switches can be used to host fog servers. All that being said, fog computing has not been proposed to replace the cloud data center, but instead can be used to complement it [10], [11].

Most future smart services are made up of multiple components, hence fog and cloud can easily cooperate to process a given service such that tasks that are resource intensive can be accommodated by the cloud whilst the less resource intensive and latency sensitive tasks can be processed by the fog [12]. We refer to such distributed infrastructures as Cloud-Fog networks in which the fog computing paradigm is organized in hierarchical layers that span the access, metro, and core domains.

This paper extends our previous proposals in [13] and [14]. In [13], we study the IoT processing placement problem, where the goal is to evaluate and compare the performance of the fog approach to the centralized processing at the cloud under two design approaches: 1) a fog network that is capacitated i.e., additional servers cannot be deployed at fog sites, and 2) an un-capacitated case whereby the number of deployed servers is unlimited at fog sites. Moreover, we looked at further scenarios such as single and multiple IoT requests, task splitting and inter-VM communication. However, the work in [13] does not take into account the mobility of the IoT devices (i.e., IoT devices are assumed to be fixed) and the heterogeneity of the communication networks, aspects that are critical for the efficient delivery of mobile IoT services. In [14], we looked at minimizing the service delay encountered by the computational offloading problem in a multi-tier edge computing system whereby multiple UAVs generate tasks for processing either on local CPUs or edge servers connected to base stations. The problem was formulated using linear programming and an algorithm was designed for real-time implementations.

Our proposed solution in this paper is of an interdisciplinary nature spanning the optimization of trajectory/path planning of UAVs and the efficient placement of services in a Cloud-Fog architecture. It is concerned with a scenario in which the decision of where to offload the UAV's computational task for processing has an impact on the trajectory the UAV takes from a predefined initial point towards a destination point. In line with the discussions above, the contributions of this paper can be summarized as follows:

- Conducted a concise review of the related works.
- Developed a MILP model to formulate a multi-objective service placement problem whereby the total power consumption of the Cloud-Fog network is jointly optimized with the trajectory of a flying UAV. In particular, the multi-objective framework includes: 1) minimizing EENPC, 2) minimizing PPC, 3) minimizing UAVTFD, and 4) minimizing UAVTPC, which includes the power consumed to transmit data and to fly the UAV from a source point to a destination point.
- Introduced a flexible mathematical model for the Cloud-Fog architecture that characterized the heterogeneity in efficiency of communication and processing resources across the access, metro, and core network layers.
- Discretized the UAV coverage environment into a square grid (which was modelled as a connected graph) and modelled the UAV trajectory on a 2-D plane between predefined set-off and destination points.
- Evaluated the proposed solution under several scenarios that required the joint orchestration of the UAV device trajectory and the multi-layered fog system. These scenarios included 1) weighted factors to emphasize and deemphasize different cost functions, 2) various CPU workloads, and 3) different antenna gains and UAV propulsion efficiency that may incentivize/disincentivize offloading data to distant macro base stations (MBSs) or pico base stations (PBSs), respectively.

The remainder of this paper is organized as follows: Section II presents a review of the related work, Section III provides in depth description of the evaluated Cloud-Fog architecture, Section IV introduces the MILP formulation of the UAV-based service placement problem. Section V presents performance evaluation and the results discussions. Section VI concludes the paper and highlights future extensions.

TABLE I
SUMMARY OF THE REVIEWED WORKS

| No. | Ref | Research Group | | Performance Metric(s) | Optimization Technique(s) | Type of Network | Scope of Optimization | |
|---|---|---|---|---|---|---|---|---|
| | | A | B | | | | UAV Trajectory | Network Resource Allocation |
| 1 | [1] | | ✓ | Mission completion time. | ML, combinatorial optimization. | N/A | ✓ | |
| 2 | [6] | ✓ | | # of served requests. | MINLP. | IoT | ✓ | |
| 3 | [15] | ✓ | | Computation rate/second. | Distributed algorithms. | SAGIN | | ✓ |
| 4 | [16] | ✓ | | # of served requests. | MINLP. | IoT/edge | ✓ | ✓ |
| 5 | [17] | ✓ | | Latency. | MILP. | IoT | ✓ | ✓ |
| 6 | [18] | ✓ | | # of UAVs deployed. | MDP, LP. | SAGIN | | ✓ |
| 7 | [19] | ✓ | | Delay. | MINLP, heuristics. | Cloud-edge | ✓ | ✓ |
| 8 | [20] | ✓ | | # of POIs. | MDP, DRL. | N/A | ✓ | ✓ |
| 9 | [21] | ✓ | | Flight time. | MINLP. | Fog | ✓ | ✓ |
| 10 | [22] | ✓ | | Processing rate. | MINLP, heuristics. | Edge | | ✓ |
| 11 | [23] | ✓ | | Delay. | MINLP, heuristics. | Ad-hoc | ✓ | ✓ |
| 12 | [24] | ✓ | | Energy efficiency, delay. | ILP, bargaining game theory. | IoT | ✓ | |
| 13 | [25] | | ✓ | Shortest path. | MILP, A-Star. | N/A | ✓ | |
| 14 | [26] | | ✓ | Shortest path. | MILP, practical. | N/A | ✓ | |
| 15 | [27] | | ✓ | # of served requests. | MILP, heuristics. | NA/A | ✓ | |
| 16 | [28] | | ✓ | Flight time. | MILP, relaxation techniques. | N/A | ✓ | |
| 17 | This work | | ✓ | Energy efficiency | MILP | Cloud-Fog | ✓ | ✓ |

## II. RELATED WORK

The rapid growth in the computing capability and inherent agility of UAVs drive the recent surge of interests in incorporating these devices into a diverse set of use cases [15]. In the literature, lots of research is conducted on UAVs, we categorize them into two main groups: A) researchers that deploy UAVs to provide infrastructures for both communication (e.g., for disaster-stricken areas or data collection from sensors based in remote locations) and computation (e.g., data processing on the edge of the network such as mobile edge computing), and B) those that do not deploy UAVs to provide an infrastructure but only deploy them to complete missions such as surveillance routines, delivery of light-weight items, land surveying, etc. This work lies in group B, since the UAV is deployed to complete a mission only and does not need to (although an interesting direction to pursue in the future) provide any infrastructure, i.e., carry communication or perform processing in the proposed Cloud Fog architecture. The work in [6] studies the deployment of UAVs as data collectors for time-constrained IoT networks in future smart cities. The problem comprises of a scenario where a single UAV must collect data from multiple IoT devices before a predefined deadline, otherwise the data will lose its value. The main goal is to find a solution that optimizes the trajectory of UAVs so that the maximum number of IoT devices are served subject to the constraint that ensures the minimum data uploaded per IoT device. Another similar comprehensive work in [16] studies the delay angle for a multi-UAV enabled edge computing platform for an IoT scenario. The UAV-enabled edge computing system provides flexible assistance to IoT devices constraint by hard deadlines. A Mixed Integer Non-Linear Programming (MINLP) model is formulated with the objective of maximizing the number of served requests through the joint optimization of trajectory and resource allocation.

In [17], the authors study UAV-mounted cloudlets to provide computational resources at the edge of the network for latency critical services (modelled using M/M/1 system) in IoT networks. An optimization framework is formulated using Mixed Integer Linear Program (MILP) with the main objective of jointly optimizing the number and positions of UAV cloudlets on a 3-D plane and task offloading decisions to provision IoT devices with latency sensitive task requirements. Due to the complexity of the optimization model, a meta-heuristic-based algorithm is proposed that achieved near-optimal performance with reduced complexity.

In [15], the authors exploit the onboard computing capability of UAVs and design a distributed task offloading scheme for multi-UAV scenarios in an edge computing environment. The UAVs collaborate to form an aerial edge computing platform referred to as 'on-the-fly' computation. Simulations results prove the effectiveness of the proposed scheme for a special case and a more general one. In another work [18], the authors investigate the computation offloading problem using MDP and linear programming models for a space-air-ground integrated network (SAGIN) . The role of the UAV here is to augment the coverage capacity of base stations and satellites for remote IoT users. In another similar work [19], the joint optimization of trajectory and resource allocation has been studied for a cloud-edge network with the objective of minimizing the maximum computation delay encountered by IoT devices.

The work in [20] proposes to jointly optimize UAV trajectory and resource allocation. The considered use case involves a UAV which is deployed to collect information from several points of interests (POIs) in a target area with limited energy and storage capacity. Onboard CPU capability and that of a ground terminal are used to provide multiple layers of processing. The problem is formulated using MDP and an optimization algorithm is proposed based on Deep Reinforcement Learning (DRL) to address the issue.

The authors in [21], using MINLP, formulate the problem of joint task allocation and UAV flight control in order to minimize the drone's flight time which is constrained by battery capacity in an environment referred to as Internet of Drones (IoD). An online algorithm is also proposed for practical scenarios. Another work in [22] looks at resource allocation and service placement in an edge computing setting supported by multiple collaborative UAVs. Through numerical results, the proposed solution demonstrated higher average processing rate in comparison to the baseline solution which did not consider the processing capability of UAVs. However, the authors did not consider the trajectory of the UAVs. Another interesting study in [23] looked at optimizing task execution latency through the joint minimization of UAV trajectory and network resource allocation in a network supported by multiple communication links. The problem is formulated in MINLP and the solution is approximated through several algorithms due to complexity issues.

Another interesting work in [24] presents a comprehensive study on value added Internet of Things (VAIoTs) whereby UAVs are equipped with various IoT payloads for different tasks in addition to their original mission. An efficient UAV selection mechanism through the joint optimization of delay and energy consumption of UAVs for IoT services is proposed. Three optimization solutions are provided, two of which use integer linear programming (ILP) formulations to optimize: 1) UAV's energy consumption, 2) UAV's operation time, and using a bargaining game to ensure 3) fair trade-off between energy and delay.

Given the stringent capacity of UAV's onboard batteries, a key challenge for researchers is ensuring that the designed solutions can jointly optimize the mission objectives together with recharging operations during long haul flights. In this direction, the work in [1] propose an automated management system in which a machine learning (ML) model is devised to estimate the energy consumption of UAVs, taking into account different real-world factors and flight scenarios. The ML model is then leveraged in the formulation of a multi-objective combinatorial optimization framework in which a UAV is required to complete a tour mission for a set of target sites in the shortest time, while simultaneously minimizing recharging duration. An approximation algorithm is also provided that can be implemented on the UAV for online flight path tracking and re-computation in dynamic environments. Extensive numerical simulations and real-world experiments prove the effectiveness of the proposed solution.

In [25], the authors conduct a comprehensive study to evaluate the performance of two widely used tools for UAV path planning: 1) Mixed Integer Linear Programs (MILP), and 2) A-Star algorithm. These two approaches are compared for efficiency in terms of time and space for a test environment that is modelled as a 2-D rectangular grid with random distribution of obstacles. The authors considered a generic use case that comprised of the goal to deliver the UAV from the starting point to the destination point using the shortest path subject to collision avoidance constraints.

In an extended work in [26], the authors and the references within make extensive use of MILP for UAV path planning. In particular, the authors model the problem of tracking icebergs using multiple UAVs, which resembles the Traveling Salesperson Problem (TSP). Simulations and practical experiments were conducted to evaluate the performance of the proposed solutions for two scenarios: 1) a single UAV tracking multiple moving icebergs, and 2) multiple UAVs tacking multiple moving ice bergs, which was shown to be superior to the first approach. In another similar contribution in [27], the authors propose efficient path planning frameworks based on MILP for a fleet of UAVs assigned to target coverage missions. Heuristics algorithms have been proposed to tackle the complexity issue of the MILP model.

The authors in [28] demonstrate the effectiveness of MILP models in tackling the UAV path planning problem. They considered different techniques such as the finite horizon approach to alleviate the complexity usually encountered with MILP models. The considered use case comprises of a scenario in which multiple UAVs start from different positions but fly to the same destination on a 2-D plane subject to zone avoidance constraints. For the reader convenience, all the reviewed works are summarized in TABLE I by their research group, performance metric(s), optimization technique, type of network and the scope of the optimization. Providing an extensive review of all the related works is out of the scope of this paper. Interested readers can refer to the following references for further useful studies related to this work [29], [30], [31], [40]–[42], [32]–[39]. While existing literature provide a significant number of UAV-based studies that describe various models to optimize UAV energy efficiency and network resource allocation in different environments. To the best of our knowledge, this is the first work that introduces a generic and modular MILP model describing the joint problem of UAV energy efficiency, processing task offloading and network resource allocation across the different domains of the end-to-end Cloud-Fog architecture.

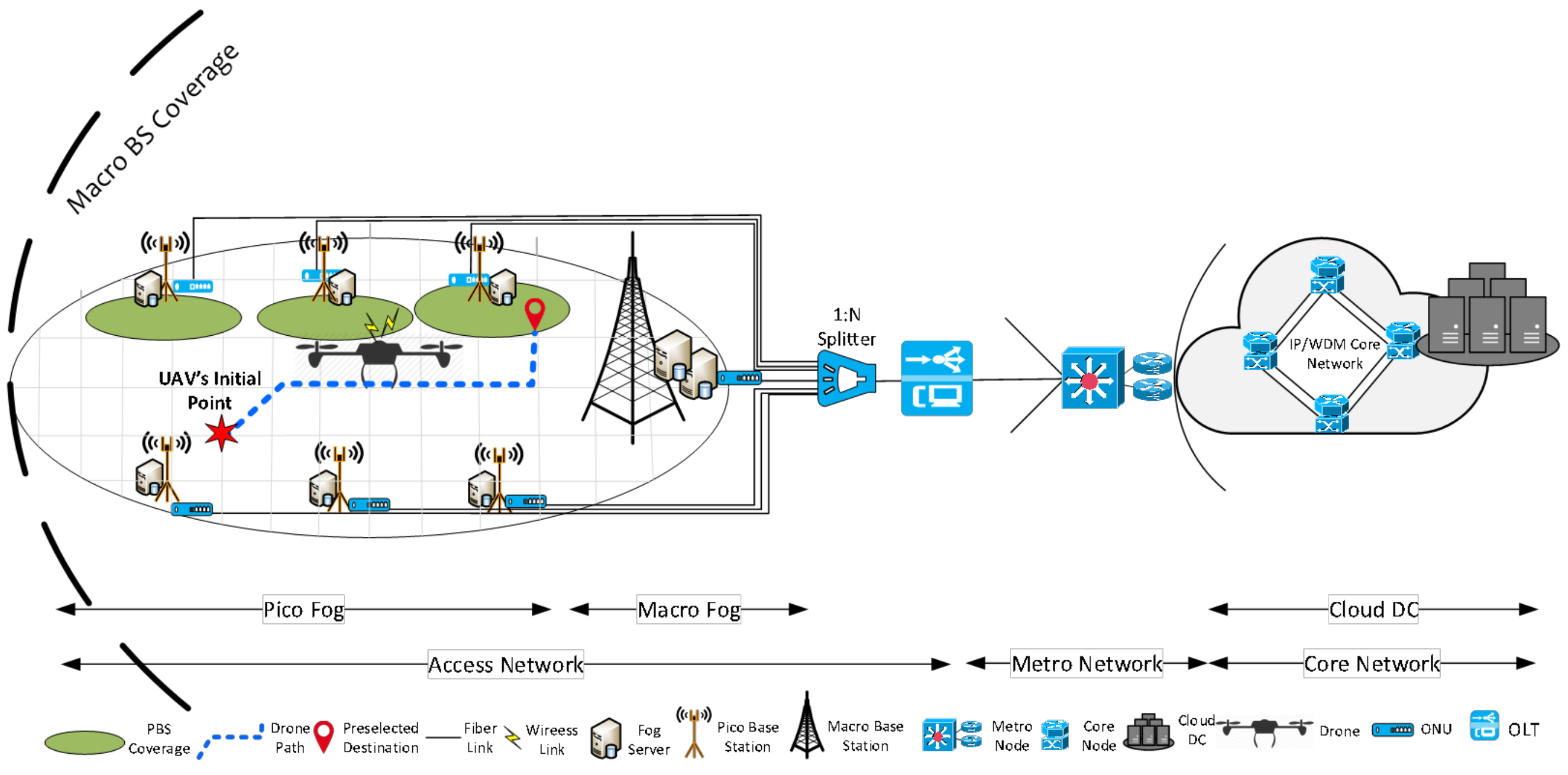

**FIGURE 1. The Proposed UAV Use Case over a Cloud-Fog Architecture.**

## III. THE EVALUATED CLOUD FOG ARCHITECURE

The proposed Cloud-Fog architecture in FIGURE 1 depicts an end-to-end deployment that spans the access, metro, and core network layers. In addition to that, the proposed architecture comprises of three processing layers, namely the Pico Fog, Macro Fog, and Cloud DC. In the following subsections, we present all layers and their details.

### A. ACCESS NETWORK

In this layer, a wired Passive Optical Network (PON) infrastructure is deployed for the back-haul due to its high bitrates and its suitability for traffic intensive applications, cost efficiency and scalability [43]. A single fiber strand can be extended from the Optical Line Terminal (OLT) and its signal can be split in the ratio of 1:N using a passive optical splitter. Optical Networking Units (ONUs) are connected to each fiber link branch to aggregate data from the end-devices. At the front-haul, wired or wireless point of access (i.e., base stations, Wi-FI APs, etc.) can be used to aggregate data from the connected end-device(s). Small and large enterprises can deploy fog units at different locations of the access network, for instance at base stations, routers, switches, or gateways [11]. The size/capacity of the fog unit can vary, depending on the site the fog is deployed i.e., fog units deployed at small cells may be less powerful compared to fog units at macro cells [44].

### B. METRO NETWORK

The metro layer consists of a high capacity Ethernet switch and a couple of edge routers that act as a gateway to the core network [45]. The Ethernet switch is typically located at locations that provide access to public clouds and is mainly used for traffic aggregation from one or more access networks. The edge routers are mainly responsible for traffic management and authentication [43]. Fog servers can be connected to metro switches and the computational resources available at metro sites can be substantially higher than those in the access network due to the number of users and services it supports [44].

### C. CORE NETWORK

The core network uses an IP over WDM network, and it consists of two layers: 1) IP Layer, and 2) and optical layer. In the IP layer, a core router is deployed at each node to aggregate network traffic from the metro routers. The optical layer is used to interconnect the core routers via optical cross connects and IP over WDM components such as EDFAs, transponders, multi/de-multiplexers, and regenerators [46]. Cloud data centers are connected to core nodes via edge routers (similar to those found in the metro layer). Due to space limitations, cloud data centers (CDCs) are designed to support a larger number of applications and services, hence clouds possess substantially more storage and computing resources in comparison to the fog [47].

## IV. MILP MODEL

This section describes the MILP model that is designed to minimize a weighted multi-objective function for the UAV's offloading and trajectory optimization problem in the Cloud Fog architecture depicted in FIGURE 1. We model the network topology as a directed connected graph which comprises of a set of nodes and a set of links connecting those nodes. Before introducing the MILP model, the sets, parameters, and decision variables are defined as follows:

| | |
|---|---|
| Sets | |
| $CN$ | Set of cloud data center(s). |
| $MN$ | Set of metro node(s). |
| $ON$ | Set of OLT node(s). |
| ONU | Set of ONU nodes. |
| $PB$ | Set of Pico base stations. |
| $MB$ | Set of Macro base stations. |
| $AP$ | Set of all access points, where $AP = PB \cup MB$. |
| $GP$ | Set of all points in a 16×6 square grid. |
| $GN_f$ | Set of neighborhood nodes of node $f \in GP$. |
| $\psi$ | Set of coverage points of PBSs. |
| $\Psi$ | Set of coverage points of MBSs. |
| $CP$ | Set of all coverage points, where $CP = \psi \cup \Psi$. |
| $PN$ | Set of processing nodes, where $PN = CN \cup ON \cup BS$. |
| $N$ | Set of all nodes in the cloud-fog architecture, where $N = PN \cup MN \cup CP$. |
| $N_m$ | Set of neighboring nodes of $m \in N$. |
| $\mathcal{S}$ | UAV's initial point, where $\mathcal{S} \subset GP$. |
| $\mathcal{D}$ | UAV's destination point, where $\mathcal{D} \subset GP$. |
| $\alpha$ | Weight to emphasize networking power consumption. |
| $\beta$ | Weight to emphasize processing power consumption. |
| $\gamma$ | Weight to emphasize UAV flight distance. |
| $\omega$ | Weight to emphasize UAV total power consumption. |
| Parameters | |
| $CPU$ | UAV's processing demand in MIPS. |
| $BR$ | UAV's bitrate demanded in Mbps. |
| $\exists$ | Energy consumed per meter by the UAV. |
| $M$ | A large enough number. |
| $\Pi_n^{(net)}$ | Maximum power consumption of network device $n \in N, n \notin CP$. |
| $\pi_n^{(net)}$ | Idle power consumption of network device $i \in N, i \notin CP$. |
| $C_n^{(net)}$ | Bit rate of network device $i \in N, i \notin CP$ in Mbps. |
| $\mathcal{E}_m^{(bit)}$ | Energy per Mbps of network device $m \in N, m \notin CP$, where $\mathcal{E}_m^{(bit)} = \frac{\Pi_m^{(net)} - \pi_m^{(net)}}{C_m^{(net)}}$. |
| $\Pi_v^{(cpu)}$ | Maximum power consumption of processing device $v \in PN$. |
| $\pi_v^{(cpu)}$ | Idle power consumption of processing device $v \in PN$. |
| $C_v^{(mips)}$ | Capacity of processing device $v \in PN$ in MIPS. |
| $\mathcal{E}_v^{(mips)}$ | Energy per bit of network device $p \in PN$, where $\mathcal{E}_v^{(mips)} = \frac{\Pi_v^{(cpu)} - \pi_v^{(cpu)}}{C_v^{(cpu)}}$. |
| $\Pi_n^{(LAN)}$ | Maximum power consumption of network device $n \in PN$ used to provide LAN between servers. |
| $\pi_n^{(LAN)}$ | Idle power consumption of network device $i \in PN$ used to provide LAN between servers. |
| $C_n^{(LAN)}$ | Bit rate of network LAN device $n \in PN$ in Mbps. |
| $\mathcal{E}_m^{(LAN)}$ | Energy per Mbps of LAN device $m \in N, m \notin CP$, where $\mathcal{E}_m^{(LAN)} = \frac{\Pi_m^{(LAN)} - \pi_m^{(LAN)}}{C_m^{(LAN)}}$. |
| $PUE_i$ | Power Usage Effectiveness of node $i \in N, i \notin CP$. |
| $IPR_n^{(net)}$ | Idle power ratio of network node $n \in N, n \notin CP$. |
| $IPR_P^{(cpu)}$ | Idle power ratio of processing node $p \in PN$. |
| $IPR_n^{(LAN)}$ | Idle power ratio of LAN equipment at processing node $p \in PN$. |
| $D_{ij}$ | Euclidean distance between grid points where $D_{ij} = \sqrt{(x_2 - x_1)^2 + (y_2 - y_1)^2}$. |
| $\mathcal{X}$ | Energy consumed per bit by the electronics of the transceiver, where $\mathcal{X} = 50nJ/bit$ [48]. |
| $\mathcal{T}$ | Energy consumed per bit by the transmission amplifier, where $\mathcal{T} = 255pJ/bit/m^2$ [48]. |
| $VS_v$ | Maximum number of servers deployed at processing location $v \in PN$. |
| Variables | |
| $\kappa_{ij}^{sf}$ | $\kappa_{ij}^{sf} = 1$ if for source $s \in \mathcal{S}$ and offloading point $f \in CP$, the drone flies over the predefined path $(i,j)$, where $i \in GP$ and $j \in GN_i$, otherwise $\kappa_{ij}^{sd} = 0$. |
| $\kappa_{ij}^{fd}$ | $\kappa_{ij}^{fd} = 1$ if for offloading point $f \in CP$ and destination point $d \in \mathcal{D}$, the drone flies over |
| | the predefined path $(i,j)$, where $i \in GP$ and $j \in GN_i$, otherwise $\kappa_{ij}^{sd} = 0$. |
| $\kappa_{ij}$ | $\kappa_{ij} = 1$, if path $(i,j)$ is used by the drone, otherwise $\kappa_{ij} = 0$. |
| $P^{fav}$ | Amount of processing aggregated by access point $a \in BS$ received from UAV at offloading point $f \in CP$ to be processed at processing node $v \in PN$. |
| $ns_v$ | Number of activated servers at processing node $v \in PN$. |
| $\lambda_{mn}^{fav}$ | Amount of traffic demand between access point $a \in BS$ and processing node $v \in PN$ aggregated from offloading point $f \in CP$ that traverses the physical link $(m,n)$, where $m \in N \backslash CP$ and $n \in N_m$. |
| $\lambda^{fav}$ | Amount of traffic demand between access point $a \in BS$ and processing node $v \in PN$ aggregated from offloading point $f \in CP$. |
| $\lambda^{fa}$ | Amount of traffic demand transmitted from offloading point $f \in GP$ towards access point $a \in BS$. |
| $\lambda_m$ | Amount of traffic aggregated on network node $m \in N \backslash CP$, where $\lambda_m = \sum_{f \in CP} \sum_{a \in GN_f} \sum_{p \in PN} \sum_{n \in N_m: m=i} \lambda_{mn}^{fav} + \sum_{f \in CP} \sum_{a \in GN_f} \sum_{p \in PN} \sum_{n \in N_m: m \neq i} \lambda_{mn}^{fav} + \sum_{f \in CP} \sum_{a \in GN_f} \sum_{p \in PN} \sum_{n \in N_m: m=v} \lambda_{mn}^{fav}$ |
| $B^{av}$ | $B^{ap} = 1$ if the aggregated traffic from AP $a \in BS$ is processed at processing node $v \in PN$, otherwise $B^{av} = 0$. |
| $B^{fav}$ | $B^{fav} = 1$ if UAV transmits from point $f \in CP$ to AP $a \in BS$ for processing at server $v \in PN$. |
| $B_m$ | $B_m = 1$ if network node $m \in N \backslash CP$ is activated, otherwise $B_m = 0$. |
| $Q_v$ | $Q_v = 1$ if processing node $v \in PN$ is activated, otherwise $Q_v = 0$ . |

The total power consumption of the Cloud-Fog network is calculated as follows:

**End-to-End Network Power Consumption (EENPC)**

$$\boldsymbol{EEN} = \sum_{\substack{m \in N: \\ m \notin CP}} \boldsymbol{PUE_m \mathcal{E}_m^{(bit)} \lambda_m} + \sum_{\substack{m \in N: \\ m \notin CP}} \boldsymbol{PUE_m \pi_m^{(net)}\, IPR_m^{(net)} B_m}. \quad (1)$$

The first term in (1) calculates the proportional power consumption of network equipment whilst the second term determines the idle power consumption.

**Processing Power Consumption (PPC)**

$$\boldsymbol{PPC} = \sum_{a \in BS} \sum_{v \in PN} \boldsymbol{PUE_v \mathcal{E}_v^{(mips)} P^{av}} + \sum_{v \in PN} \boldsymbol{PUE_v \pi_v^{(CPU)}\, IPR_v^{(cpu)} ns_v} + \sum_{v \in PN} \boldsymbol{PUE_v \mathcal{E}_v^{(LAN)}\, \lambda_v} + \sum_{v \in PN} \boldsymbol{PUE_v \pi_v^{(LAN)}\, IPR_v^{(LAN)} Q_v}. \quad (2)$$

The first term in (2) calculates the proportional power consumption of processing servers whilst the second term determines the idle power consumption. The third and fourth terms calculate the proportional and idle power consumption of server LAN equipment.

**The UAV's Total Power Consumption (UAVTPC)**

$$\boldsymbol{UAVTPC} = \sum_{f \in OP} \sum_{a \in GN_f} \boldsymbol{\mathcal{X} \lambda^{fa}} + \sum_{f \in OP} \sum_{\substack{a \in GN_f: \\ a \in MBS}} \boldsymbol{\mathcal{T}(D_{fa})^2\, \lambda^{fa}} + \sum_{f \in OP} \sum_{\substack{a \in GN_f: \\ a \in PBS}} \boldsymbol{\mathcal{T}(D_{fa})^2\, \lambda^{fa}} + \sum_{i \in GP} \sum_{j \in GN_i} \boldsymbol{\exists D_{ij} \kappa_{ij}} \quad (3)$$

The first term in (3) is used to determine the power consumption due of radio electronics onboard the UAV, the second term calculates the total power consumption due to transmission of data whilst the third term calculates the power consumption due to flying.

**UAV's Total Flight Distance (UAVTDC)**

$$\boldsymbol{UAVTDC} = \sum_{i \in GP} \sum_{j \in GN_i} \boldsymbol{D_{ij} \kappa_{ij}}. \quad (4)$$

In (4), the total flight distance travelled is calculated.

**Objective Function: Minimize**

$$\alpha EENPC \; + \beta PPC + \gamma UAVTDC + \omega UAVTPC. \quad (5)$$

which is a weighted sum of the End-to-End Network Power Consumption, Processing power consumption, UAV total power consumption and UAV total flight distance.

**Subject to:**

$$\sum_{j \in GN_i} \kappa_{ij}^{sf} - \sum_{j \in GN_i} \kappa_{ji}^{sf} = \begin{cases} 1 & i = s \\ -1 & i = p \\ 0 & otherwise \end{cases} \quad (6)$$

$\forall i \in GP, s \in S, f \in CP$

Constraint (6) ensures the flow of the UAV is preserved between the source and the offloading point.

$$\sum_{j \in GN_i} \kappa_{ij}^{fd} - \sum_{j \in GN_i} \kappa_{ji}^{fd} = \begin{cases} OP_f & i = s \\ -OP_f & i = p \\ 0 & otherwise \end{cases} \quad (7)$$

$\forall i \in GP, f \in CP, d \in \mathcal{D}$

Constraint (7) ensures the flow of the UAV is preserved between the offloading point and the destination.

$$\sum_{n \in N_m} \lambda_{mn}^{fav} - \sum_{n \in N_m} \lambda_{nm}^{fav} \begin{cases} \lambda^{fav} & i = s \\ -\lambda^{fav} & i = p \\ 0 & otherwise \end{cases} \quad (8)$$

$\forall m \in N \backslash CP, a \in BS, v \in PN$

Constraint (8) ensures that traffic flow is preserved in the network.

$$\sum_{f \in CP} \sum_{a \in N_f} \sum_{v \in PN} \lambda^{fav} = BR \quad (9)$$

Constraint (9) ensures that the traffic demand is met.

$$p^{av} = CPUB^{av} \quad (10)$$

$\forall a \in BS, v \in PN.$

Constraint (10) ensures that the CPU demand is met.

$$\sum_{a \in GN_f} \lambda^{fap} \geq B^{fav} \quad (11)$$

$\forall f \in CP$

$$\sum_{a \in GN_f} \lambda^{fap} \leq MB^{fav} \quad (12)$$

$\forall f \in CP$

Constraint (11) and constraint (12) produce the binary indicator $B^{fav}$.

$$\sum_{f \in CP} \sum_{a \in GN_f} \sum_{v \in PN} B^{fav} \leq 1 \quad (13)$$

Constraint (13) ensures a single offloading point is chosen together with a single processing location.

$$\sum_{f \in CP} \sum_{a \in GN_f} \sum_{v \in PN} \lambda_{mn}^{fav} \geq B_{mn} \quad (14)$$

$\forall m \in N, n \in N_m$

$$\sum_{f \in CP} \sum_{a \in GN_f} \sum_{v \in PN} \lambda_{mn}^{fav} \leq MB_{mn} \quad (15)$$

$\forall m \in N, n \in N_m$

Constraint (14) and constraint (15) determine the network link activation binary indicator.

$$\lambda_m \geq B_m \quad (16)$$

$\forall m \in N \backslash CP$

$$\lambda_m \leq MB_m \quad (17)$$

$\forall m \in N \backslash CP$

Constraint (16) and constraint (17) determine the network node activation binary indicator.

$$ns_v \geq \frac{\sum_{f \in CP} \sum_{a \in GN_f} p^{fav}}{C_v^{(mips)}} \quad (18)$$

$\forall v \in PN$

Constraint (18) determines the number of activated servers at processing location $v \in PN$.

$$ns_v \leq VS_v \quad (19)$$

$\forall v \in PN$

Constraint (19) ensures that the number of activated servers at processing location $v \in PN$ does not exceed the maximum number of servers deployed at that processing location.

$$B_{mn} = 0 \quad (20)$$

$\forall m \in OLT, n \in N_m : n \in ONU$

$$B_{mn} = 0 \quad (21)$$

$\forall m \in AP, n \in N_m : n \in CP$

Constraint (20) and (21) ensure that traffic only flows in the uplink direction.

$$\sum_{a \in AP} p^{av} \geq Q_v \quad (22)$$

$\forall v \in PN$

$$\sum_{a \in AP} p^{av} \leq MQ_v \quad (23)$$

$\forall v \in PN$

Constraint (22) and constraint (23) determine the value of the binary indicator $Q_v$.

## V. PERFORMANCE **EVALUATION**

In this section, we present the results of the optimization framework for illustrative UAV services over a Cloud Fog network infrastructure. The optimization framework is solved using IBM's commercial solver CPLEX over a PC with Intel i7 CPU and a RAM of 16GB. We analyze the computational offloading and UAV's trajectory planning problem via a weighted multi-objective function in which the considered cost functions are either emphasized or de-emphasized. These cost functions include the end-to-end network power consumption, processing power consumption, UAV's flight distance, and UAV's total power consumption. The considered optimization model in this work is centralized in nature and the solution to the problem is determined offline. This requires global knowledge about the Cloud Fog infrastructure as a priori such as available processing resources and network paths. However, the task placement and trajectory planning approach is dynamic as the model can deploy services or replace services (i.e., from one server to the next) as UAV workload changes over time. The readers can refer to the work in [49] for an overview of different solution frameworks for service placement in fog environments.

### D. CLOUD FOG SETUP

We consider a hierarchical Cloud Fog architecture comprising of three main layers of processing: i) a Cloud DC layer representing the most energy efficient processing node, and ii) a fog layer, in which fog nodes are organized into 2 tiers, i.e., a Macro Fog node representing the second most energy efficient processing node and a Pico Fog node representing the least energy efficient processing node. This is a valid assumption since cloud data centers due to the type of equipment deployed (i.e., servers), they offer higher resource consolidation and multiplexing gains [7]. Hence, processing efficiency decreases as we get closer to the end-device layer. As for the networking fabric, we consider a heterogenous networking environment in which two types of access points (APs) exist, namely, a Pico Base Station (PBS) and a Macro Base Station (MBS). We consider a Wi-Fi AP with a coverage of 25 meters to represent the Pico cells and an LTE base station that covers the entire deployment region. The considered MBS is a 3-sector $2 \times 2$ MIMO that is usually deployed in urban areas [50]. In total, we have considered 9 PBS and 1 MBS, each of which is connected to an ONU. The ONUs are in turn connected via passive splitters to a single OLT device (typically located in the central office (CO)). In the metro, we consider a single metro switch (MSW) and multiple edge router ports (MRPs) for redundancy purposes. Finally, we consider a single core node (CN) to represent the backbone network based on the assumption that the cloud data center is a single hop from the core node. The power consumption of the different elements within the core node and metro node has been estimated by summing the different elements in those nodes. Also the average distance (which is used to calculate the number of EDFAs [46]) between two core nodes is assumed to be 509km, estimated using google maps for the AT&T topology. We have also accounted for the Local Area Network (LAN) equipment inside processing nodes by considering fog switches and fog routers. The LAN is needed to stablish communication between several processing servers. We have also considered the power consumption due to external factors such as Power Usage Effectiveness (PUE) to measure the power efficiency of the network or processing nodes. PUE is the ratio of the power consumed for IT (ie networking and processing) to the total power consumed by the a facility (i.e., power consumption of IT plus cooling, lighting, and ventilation) [43]. Based on [55], we assume that the cloud node has the best PUE value of 1, the metro and MBS node have a PUE value of 1.12, whilst the PUE value of PBS nodes is distributed uniformly between 1.3 – 1.5.

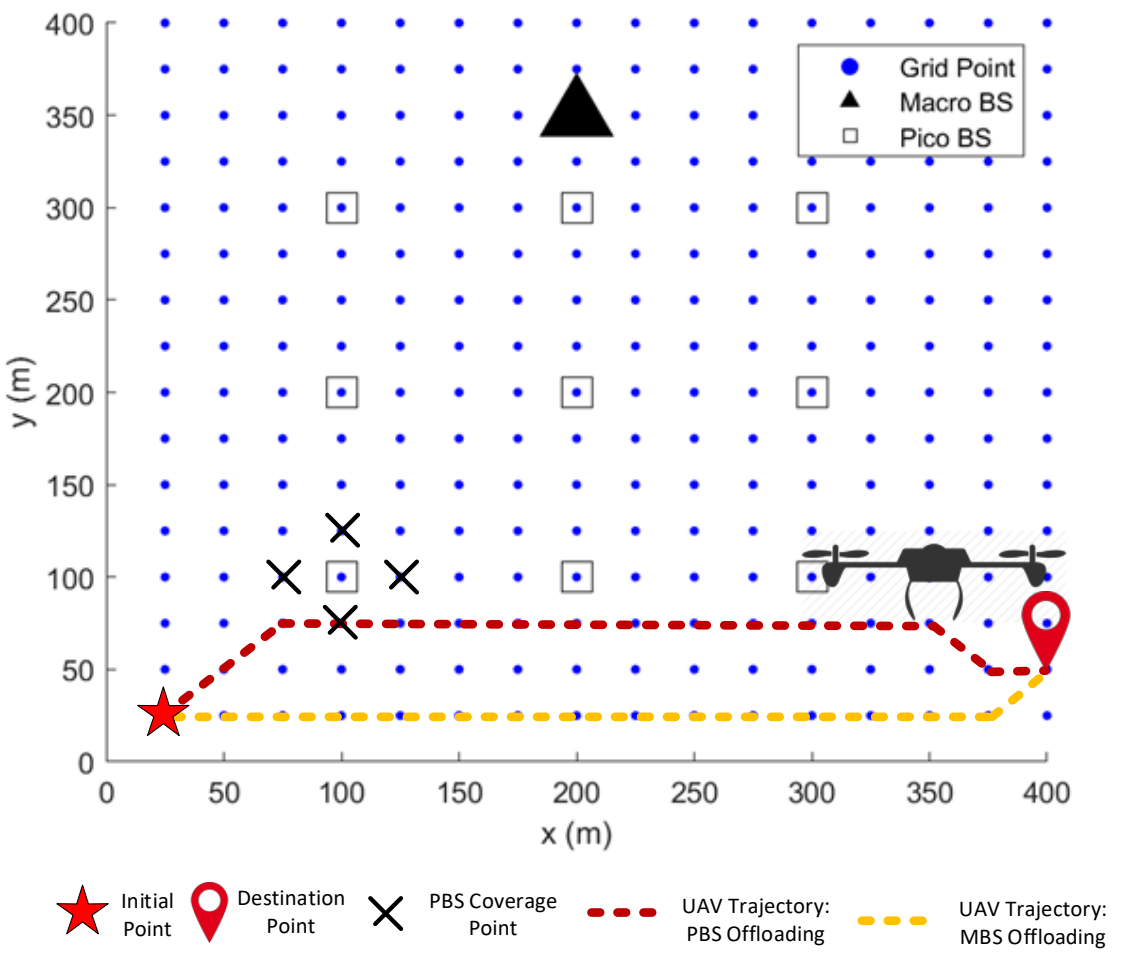

**FIGURE 2. The illustration of UAV's trajectory over a $400m^2$ obstacle free region.**

TABLE II
NETWORK PARAMETERS

| Device | Max (W) | Idle (W) | Bitrate | Efficiency |
|---|---|---|---|---|
| PBS | 21[51] | 19 | 300Mbps | 433mW/Mbps[51] |
| MBS | 528[50] | 333[50] | 72Mbps | 2.7W/Mbps |
| ONU | 15 [13] | 9 | 10Gbps | 600$\mu$W/Mbps |
| OLT | 1940 [52] | 60[52] | 8600Gbps | 219$\mu$W/Mbps |
| Metro Router Port | 30 [13] | 27 [13] | 40Gbps | 75$\mu$W/Mbps |
| Metro Switch | 470 [13] | 423 | 600Gbps | 78$\mu$W/Mbps |
| Core Node | 955 [13] | 859 | 40Gbps | 2.4mW/Mbps |
| Fog Switch | 210 [13] | 189 | 600Gbps | 35uW/Mbps |
| Fog Router Port | 13 [13] | 12 | 40Gbps | 25uW/Mbps |

TABLE III
SERVER PARAMETERS

| Server | Max(W) | Idle(W) | MIPS | Efficiency |
|---|---|---|---|---|
| Pico Fog | 180 [53] | 108 | 10768 [54] | 6.68mW/MIPS |
| Macro Fog | 450 [5] | 270 | 73193 [54] | 2.46mW /MIPS |
| Cloud DC | 495 [4] | 297 | 293415 [54] | 675uW/MIPS |

TABLE IV
UAV PARAMETERS (DJI MAVIC PRO)

It is worth noting that the PUE at both processing layer and networking layer is identical if the two nodes are collocated. The networking parameters are summarized in TABLE II. In TABLE III, we summarize the parameters of the processing nodes. The processing capacity of the servers in MIPS has been estimated using publicly available benchmarking forum in [54]. In a similar fashion to the way we assigned PUE in the Cloud Fog architecture, we assume that the most energy efficient server is deployed at the Cloud DC, followed by the Macro Fog and then the Pico Fog.

### E. NETWORK DEPLOYMENT & USECASE

The UAV environment is decomposed into a square grid spanning an area of $400m^2$, as shown in FIGURE 2. We have assumed that, on the 2-D plane the UAV moves in steps of 25 meters, hence in total, we have 256 points on the $16 \times 16$ grid. The distances between any two grid points are calculated as the Euclidean distance. We consider a generic use case in which a UAV is required to travel between two given points, i.e., an initial point and destination point. It is assumed that the UAV has generated a monolithic service [49] that comprises of a single component only. This is a valid assumption as the UAV is assumed to have collected some ground data (i.e., high-definition video files) at a previous mission for which a computational task has been generated, which needs to be offloaded prior to arriving at the next destination point. The UAV's trajectory on the 2-D plane involves a single step movement in the X and Y coordinates. For instance, a single step is taken either in the vertical, horizontal, or diagonal direction. This method is particularly useful as it reduces the complexity of the MILP model [25]. Partly motivated by the work in [48] and [56], we assume that the ONU devices connected to the PBSs are uniformly placed at regular intervals (every 100 meters) whilst the ONU that is connected to the MBS is placed at the top-center of the grid, as can be seen in FIGURE 2. As illustrated, the trajectory of the UAV depends on the offloading decision i.e., offloading via the PBS or offloading via the MBS. To offload via the PBS, the UAV must visit one of the coverage points of the chosen PBS whilst with the MBS, this is not required as it can cover the entire region, thus the UAV offloads its data shortly after the initial point.

### F. POWER CONSUMPTION MODEL & DATA

In this work, the total power consumption of different equipment is modelled as the sum of two parts. The first part describes a proportional power consumption, which is a function of the total workload (processing or traffic) and the second part is the idle power consumption, which is consumed as soon as the equipment is activated. This is a widely used methodology in the literature [57] and is consistent with our previous contributions cited in Section I, power consumption is a function of the volume of the workload (processing or traffic) and the idle power is consumed as soon as the equipment is activated. The power consumption values in

$$TPC = \left(\frac{MaxPC - IdlePC}{CAP} Load\right) + IdlePC \qquad (24)$$

| | |
|---|---|
| Battery Capacity | 157183 Joules |
| Voltage | 11.4 V |
| Maximum Flight Time | 21 minutes |
| Maximum Power Consumption | ~125W |
| Maximum Flight Distance | 13km |
| Power Per Meter | ~9.6mW |

TABLE II – TABLE IV have mostly been cited from equipment datasheets where possible, otherwise we have made assumptions based on the literature for determining the other parameters. It is worthy of mention that precisely modelling the energy efficiency of the equipment is a cumbersome task due to the limited information disclosed by manufacturers [7]. Accordingly, the idle power consumption of highly shared networking equipment (i.e., routers and switches) is assumed to be 90% of the maximum power consumption based on the work in [50]. As for the processing devices, we estimate the idle power consumption of servers to be 60% of the servers' maximum power consumption based on [58]. It is important to note that for the core network, we take an aggregate power consumption by taking the sum of the power consumption of all the elements such as core router ports, EDFAs, optical switches, transponders, etc. This is due to modelling simplicity and the fact that only a single core node was considered. Thus, the total power consumed by networking or processing equipment can be calculated using (24), where TPC is the total power consumption, MaxP is the peak power consumption, IdlePC is the idle power consumption, CAP is the capacity of the equipment (bps or MIPS) and the Load is traffic or processing load:

It is important to note that the Cloud Fog architecture under consideration comprises of a single access network with 10 ONU devices aggregating data from a single UAV and then as we go deeper into the network, we have highly shared equipment in the metro and core nodes. This makes the considered UAV application out of proportion in comparison to the Cloud Fog deployment. To address this issue, we have refined the approach, where the idle power consumption considered throughout the network is proportional to the amount of workload (processing or traffic) as it would be unjust to ascribe the total idle power consumption to the UAV application alone.

As for the UAV's energy consumption model, three main sources are considered in the literature [24]: 1) energy consumed by the UAV during flight activities, 2) energy consumed to perform a task (e.g., sensing or video streaming), and 3) the energy needed for data communication. In this work, we only consider sources 1) and 3). In addition, flight energy consumption is comprised of different maneuvers such as ascending, descending, flying horizontally, and hovering [31]. For simplicity purposes, we assume the flight power consumptions includes all the aforementioned maneuvers. The UAV's power consumption is also affected by the other factors such as speed, payload, and route distance. We consider a representative drone device and we cite the default key parameters from its manufacturer's website in [59]. Assuming 21 minutes of overall flight time on a full battery that has a capacity of $157183J$, we estimate the total power consumption of the UAV to be approximately 125W (hence $\frac{157183J}{21\times60} \cong 125W$). Then, we divide the total power consumption by the maximum flight distance permitted on a full battery to determine the amount of power consumed per meter. This method simplifies the modelling complexity however, interested reader can refer to [27], [60], and [61] for detailed power consumptions models based on empirical studies conducted on different UAV maneuvers such as horizontal, vertical, ascending and descending as well as the impact of payloads. For the communication induced power consumption of the UAV i.e., power consumed for sending and receiving data, we take a similar approach to the work in [62], which is based on the Friis free-space equation defined as follows:

$$P_{tx} = \left(E_e + \frac{\epsilon d^2}{G}\right)\lambda \tag{25}$$

where $P_{tx}$ is the transmitter power consumption, $E_e$ is the energy consumed by the radio electronics, $\epsilon$ is the energy consumed by the transmit amplifier, G is the gain of the antenna, and $\lambda$ is the transmitted data in bps.

### G. TASK OFFLOADING FLOW PROCESS

In this work, it is assumed that the task offloading and task allocation flow process follows the steps below:

1. A task is generated by the UAV and through periodic control signals, a centralized controller [24] that has full knowledge of the network's resources is synchronized. This includes complete knowledge on the UAV's mission, the required volume of processing workload and the associated data rate.
2. The connected BS transmits acknowledgement signals to the UAV responsible for generating the task(s).

These two steps are not considered in the model as they are considered control signals, therefore, generating small traffic volumes that consume negligible power in comparison to raw data transmissions. Thus, only uplink data transmission is considered, similar to the work in [51].

3. The UAV offloads the collected data via PBSs and MBs for processing in the Cloud Fog network. As this step carries the raw data, it will impact the task allocation and UAV's trajectory. Therefore, it is considered as the main element of the model.
4. The output (e.g., extracted knowledge) as a result of the processing the collected data is sent back from the BS in the downlink direction to the UAV's control center.

Since the extracted knowledge carries negligible amount of data [63], this step is not included in the model.

### H. RESULTS AND DISCUSSIONS

#### 1) TEST CASE #ONE

This scenario comprises of a Cloud-Fog network that has proportional – cost – and – un-capacitated – servers. This assumes that the idle power consumption of both networking and processing equipment is proportional to the application workload in terms of traffic and CPU, respectively. We also assume that task splitting is not permitted in the model. In addition, we set the gain of the UAV and PBSs to 1 and set the gain of the MBS much higher (typical MBSs have a gain of 100). We assume the UAV's coordinates (x, y) for the initial point on the grid is set to (1, 1) whilst the coordinates of the destination point are set to (32, 2). Figure 3(a) shows the network power consumption against CPU workloads under different objective functions. As for the workload, we assume different CPU demand intensities ranging from 10% to 100% of the maximum capacity of a Pico Fog server whilst we keep the data rate constant at 10Mbps.

As was expected when the network power consumption is given precedence over the other cost functions, the network power savings are substantial. In this case, the UAV offloads its data via the PBSs, which are much more energy efficient than the MBSs. With large values of $\beta$ and $\omega$, the network power consumption is comparable to the case when $\alpha$ is large because both choose to offload via the PBSs, albeit choosing PBSs with slightly higher PUE value. However, the worst-case scenario for the network is when $\gamma$ is given importance i.e., to reduce the distance to the destination, hence the UAV offloads its data via the MBS due to its greater coverage and the optimization results are indifferent about where the data is processed, hence processing takes place in the Cloud DC and the number of networking elements activated is much higher than if it was processed at the fog layer. Also, transmitting data to the MBS means transmitting at a much longer distance, which also adds more onto the total network power consumption. It is worth noting that the majority of the end-to-end network power consumption is ascribed to the idle power consumption of all the elements that are activated to get to the Cloud DC. This should concern network designers as we can observe from this result that even a small idle power consumption, has a significant impact due to the number of hops between end-device and the Cloud DC.

Figure 3(b) shows the processing power consumption. In this work, the cloud data center is the most energy efficient location for processing due to its hyper scale deployment and the use of highly efficient servers. The worst-case scenario is when the EENPC cost function is given importance ($\alpha \gg$ $(\beta + \gamma + \omega)$, which forces the model to offload data to the PBSs, hence processing takes place on one of the Pico Fogs which results in high incurred processing power consumption. Therefore, we can conclude that the most efficient approach is the one that minimizes both segments simultaneously. The power consumed under the remaining scenarios ($\gamma \gg (\alpha + \beta + \omega)$ and $\omega \gg (\alpha + \beta + \gamma)$) is comparable due to the fact that in both scenarios processing workload is offloaded to the cloud data center.

Figure 3(c) shows the UAV's total flight distance (UAVTFD) between the source and destination, in meters. The distance is optimized when $\gamma \gg (\alpha + \beta + \omega)$. The worst-case scenario is when EENPC function is emphasized because the UAV is forced to take a longer route in order to be able to transmit the collected data under the coverage of one of the energy efficient PBS, hence PBS's coverage area is considerably smaller (25 meters) than the MBS. However, under the PPC and UAVTPC solutions, the UAV still takes slightly longer routes due to the fact that with the PPC the UAV offloads to a PBS that has the best PUE (i.e., network and processing nodes are associated with random PUE values, some with better PUE may be further away whilst the worst ones may be in close proximity) and with UAVTPC, the UAV chooses any of the PBSs randomly as it does not care about PUE. An observation to make here is that minimizing the flight distance does not necessarily mean minimizing the UAV's total power consumption unless the propulsion efficiency of the UAV is considerably bad.

Figure 3(d) shows the total power consumed by the UAV (UAVTPC) for data communication and mobility. As was expected, the $\gamma \gg (\alpha + \beta + \omega)$ approach which was focused around minimizing the total flight distance (UAVTFD), produced the highest power consumption as in this case emphasis was put on the shortest distance forcing the UAV to offload via the MBS due to its greater coverage range and avoiding the extra number of meters in the trip to offload via a PBS. It is worth noting that minimizing the total flight distance could minimize the UAV's total power consumption, however this depends on the proportion of the number of meters avoided to offload via the PBSs compared with the total number of meters in the whole journey. In this direction, we vary the gain of the MBS to evaluate its impact on the UAV's offloading behavior under the scenario where $\omega \gg (\alpha + \beta + \gamma)$, i.e., minimizing the UAVTPC function. The results in Figure 4(a), Figure 4(b) and Figure 4(c) show the total flight distance, UAV's total power consumption and a breakdown into the power consumed for data transmission and power consumed during a flight routine, respectively. As shown Figure 4(a), when the gain of the MBS is improved (Gm=200), the UAV takes a shorter route to the destination, i.e., several hops are avoided as the UAV can offload from anywhere on the grid via the MBS (in this work, due to complexity reasons the UAV is assumed to offload shortly after the initial point). In all the results shown in FIGURE 3, we assumed that the MBS's antenna had a gain of 100 (i.e., 20dB), thus for the scenario where $\omega \gg (\alpha + \beta + \gamma)$, we observed that the model was incentivized to offload via the PBSs and as result trade off the power consumed by the UAV for traveling with the power consumed by the UAV for communication. However, for the same scenarios, we re-run the model but this time we varied the gain of the MBS's antenna from 100 – 200 in steps of 50 to represent scenarios

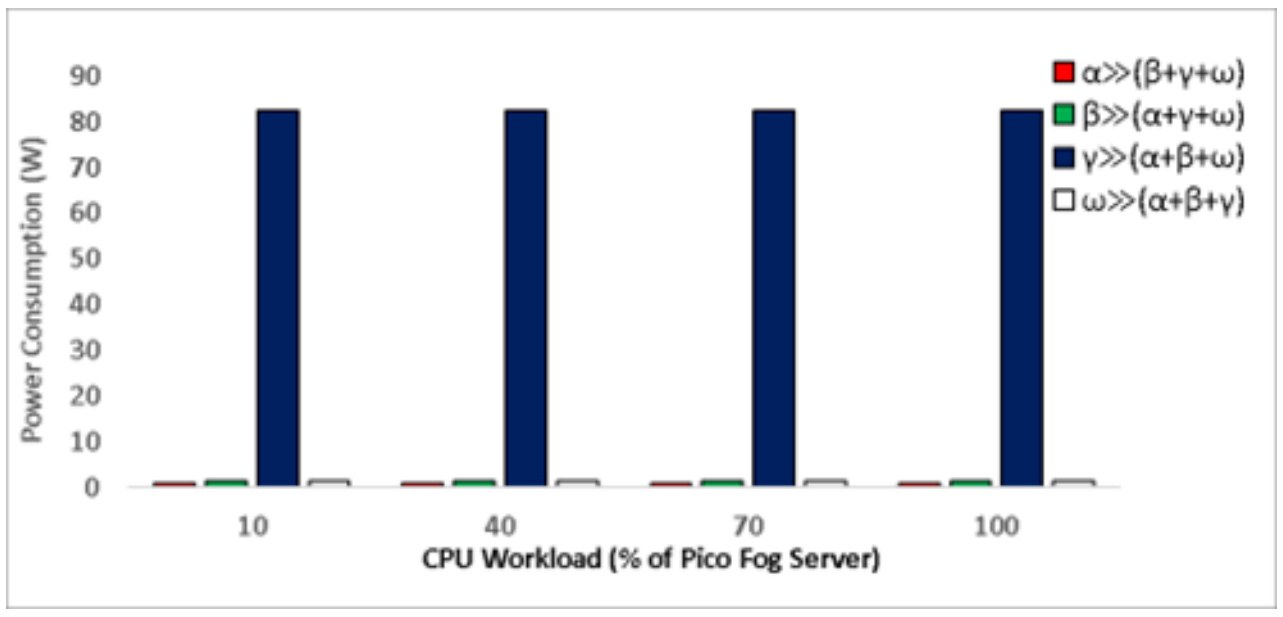

(a)

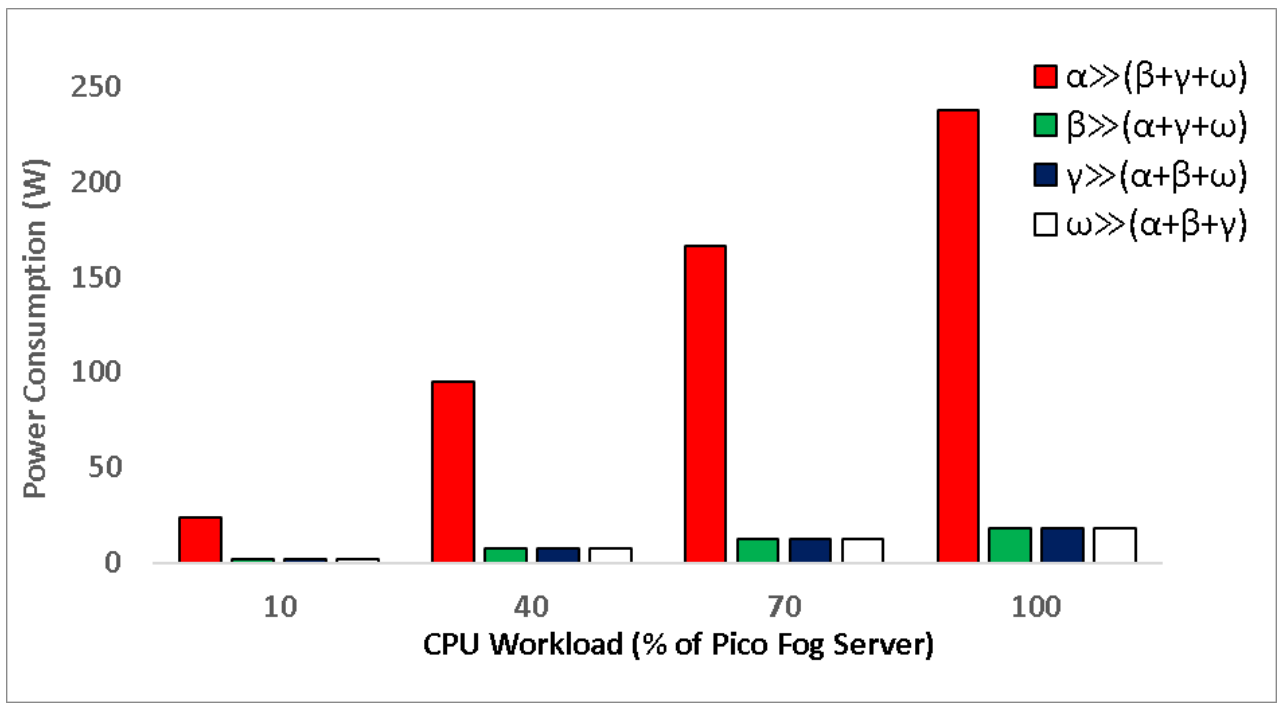

(b)

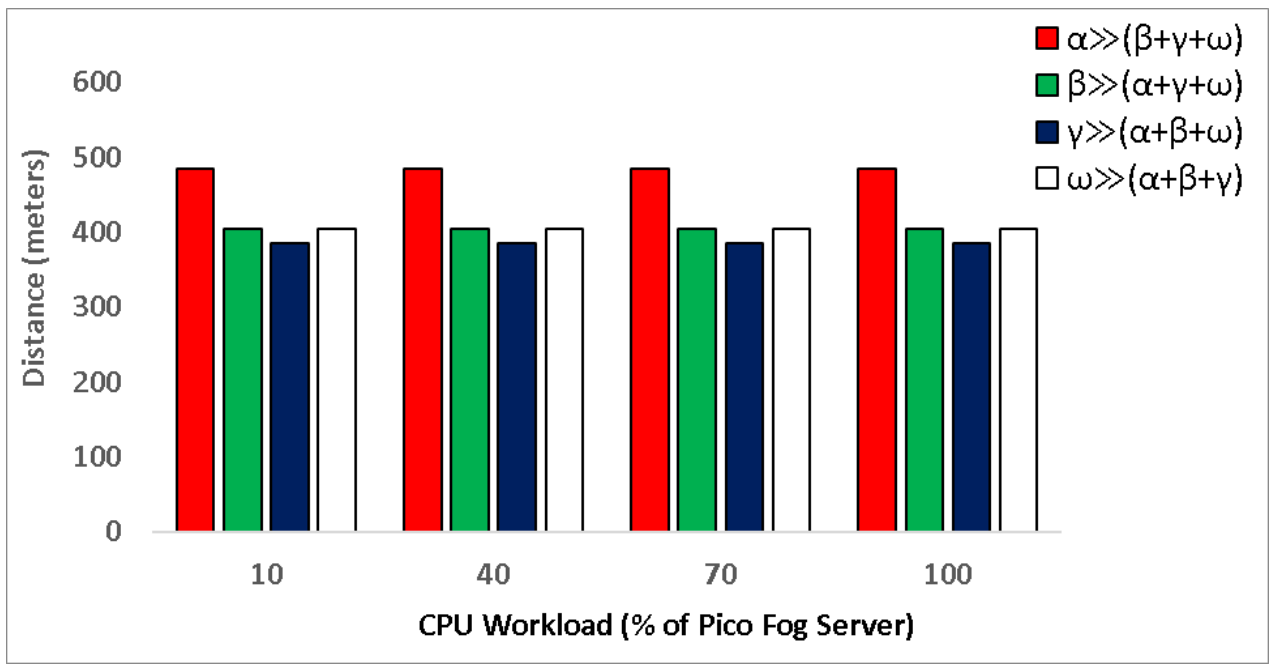

(c)

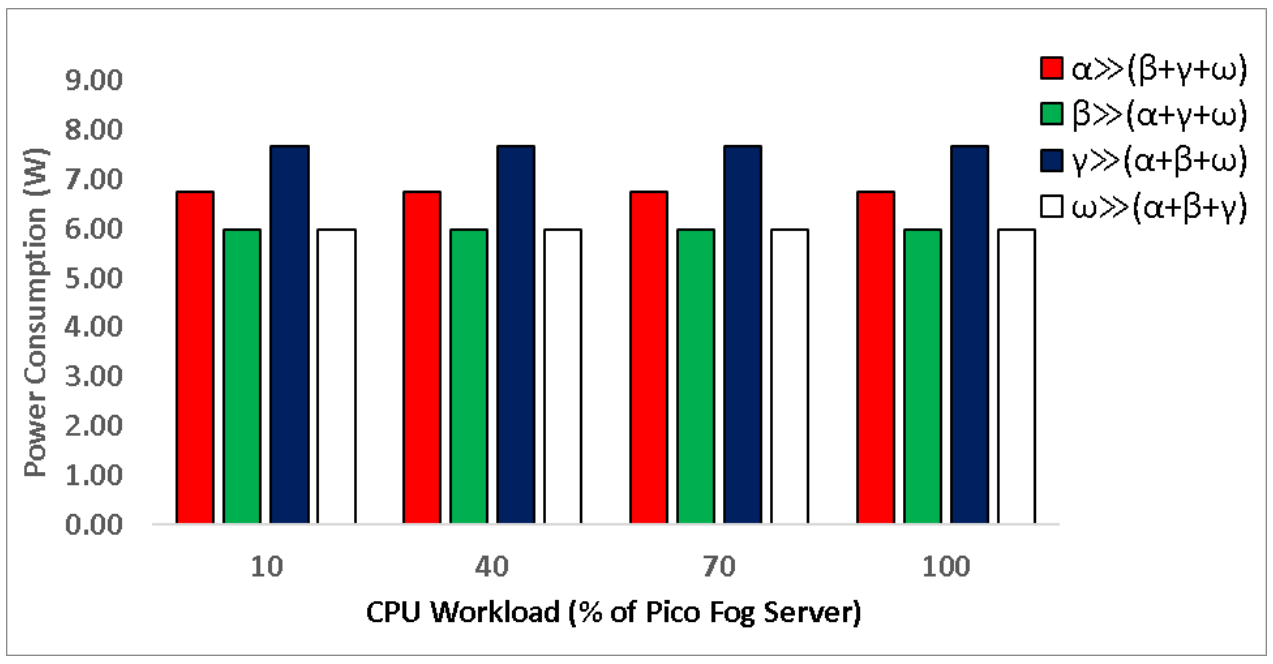

(d)

**FIGURE 3. Power consumption versus CPU workload under different objective functions to minimize only: (a) end-to-end network power consumption of the wired infrastructure (EENPC), (b) processing power**

**consumption (PPC), (c) UAV's total flight distance (UAVTFD), and (d) UAV's total power consumption (UAVTPC).**

where the MBS's efficiency is increased gradually. The results in FIGURE 4 show the UAV's total flight distance and total power consumption as well as a breakdown of the total power consumption into that of transmission and flight. In FIGURE 4(a), the total flight distance is reduced when the gain of the MBS is set to 200, as was expected since data transmission with the MBS enables the UAV to travel in an almost straight line from the initial point to the destination, hence less propulsion power is consumed due to a shorter traveled distance. In total, 0.05% of UAV's power consumption saving was achieved due to travelling shorter distance than the scenario where the offloading took place via one of the PBSs (Gm=100 and Gm=150). We can conclude from this result that the trajectory pattern of the UAV has negligible impact on the UAV's total power consumption if the UAV device has an efficient value for the energy consumed per meter as shown in FIGURE 4(b). As for the transmission induced power consumption, as can be seen in FIGURE 4(c), the power consumption is increased by 0.08% compared to the scenario where the UAV offloaded its data to the PBS from a 25m distance.

Due to the deployment scenario and the considered test case, PBSs are placed relatively close to the UAV, hence saving the extra propulsion power needed to get to the PBSs has negligible impact on the UAV's total power consumption. Also, since the considered gain values of the MBS are relatively high, any power consumed by communicating over a long distance (i.e., offloading to MBS from 381m) compared to communicating over a shorter distance (i.e., offloading to PBSs from 25m) will still be relatively comparable, hence this translates to smaller power savings as seen in FIGURE 4(b). We can conclude that the UAV's power consumption is a function of both the distance travelled, UAV's propulsion efficiency, number of bits transmitted, the range of communication, gain of the BS, UAV's initial and destination points. In line with the above discussion, following subsections will explore further cases.

### 2) TEST CASE #TWO

In this test case, we evaluate the power consumption of the UAV by varying its propulsion efficiency to see at what point it is favorable to offload via a distant PBS in comparison to offloading via an energy efficient MBS that has an antenna gain of 200. We aim to represent a scenario in which Pico Fogs are not always available due to their capacity limitations and the number of users they can support; hence we assume that in order for the UAV to be able to offload via the PBS, it must travel a further distance to get to it. On the other hand, an energy efficient MBS is assumed to be always available. We estimate the UAV's propulsion efficiency by varying the distance it can travel using the default battery cited in TABLE IV. We simply take a proportion of the distance; hence higher UPE values imply better propulsion efficiency. In FIGURE 5, we present the results under two options: 1) a full offload option and 2) limited offload option with the first options, the model is free to offload via a distant PBS or the MBS. However, with the limited option, the UAV is forced to offload via the distant PBS only. This is a valid assumption because it may not always be possible to always have access to ubiquitous services. As can be seen, when the UAV's propulsion efficiency (UPE) is at its worst (10%), offloading via the MBS is the best choice and a maximum power saving of 34% can be achieved. This is a plausible outcome as the power consumed during the flight routine to get to the distant PBS is far more superior to the power consumed by transmitting over a long distance to the MBS, as shown in FIGURE 6(b). We can conclude that even for a very highly efficient UPE, offloading via an MBS is much better than flying longer distances to get to a low power PBS, hence power savings of about 13% is achieved even at UPE=100% as shown in FIGURE 5.

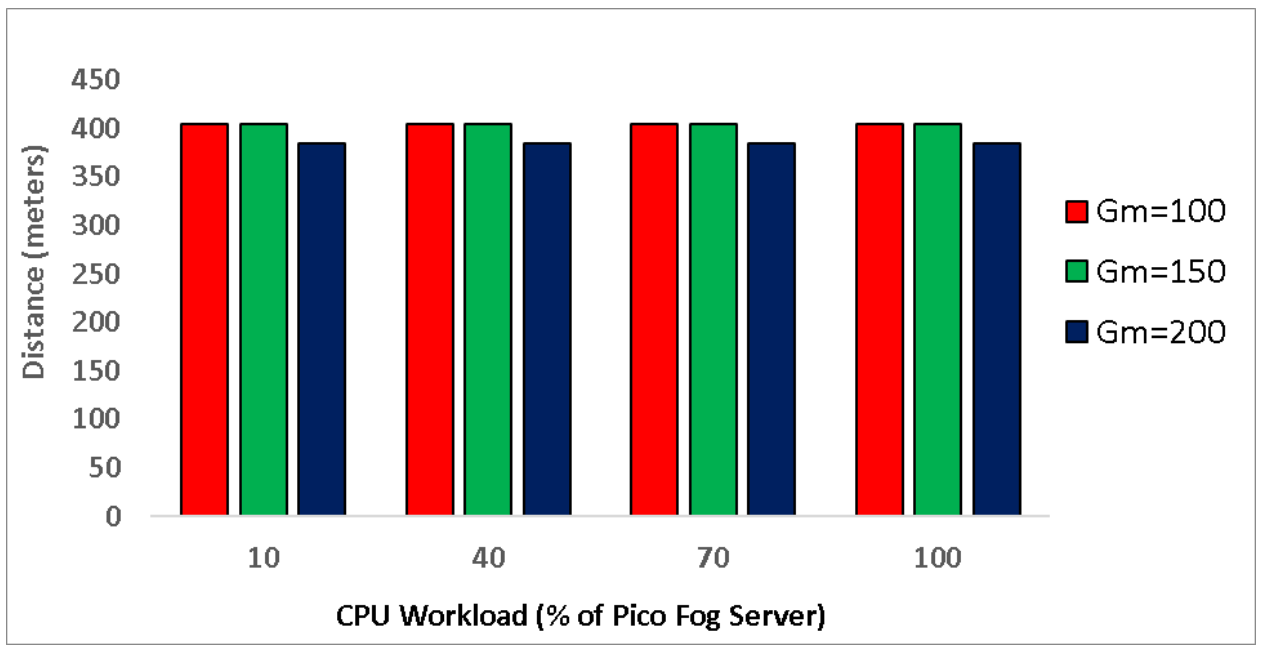

(a)

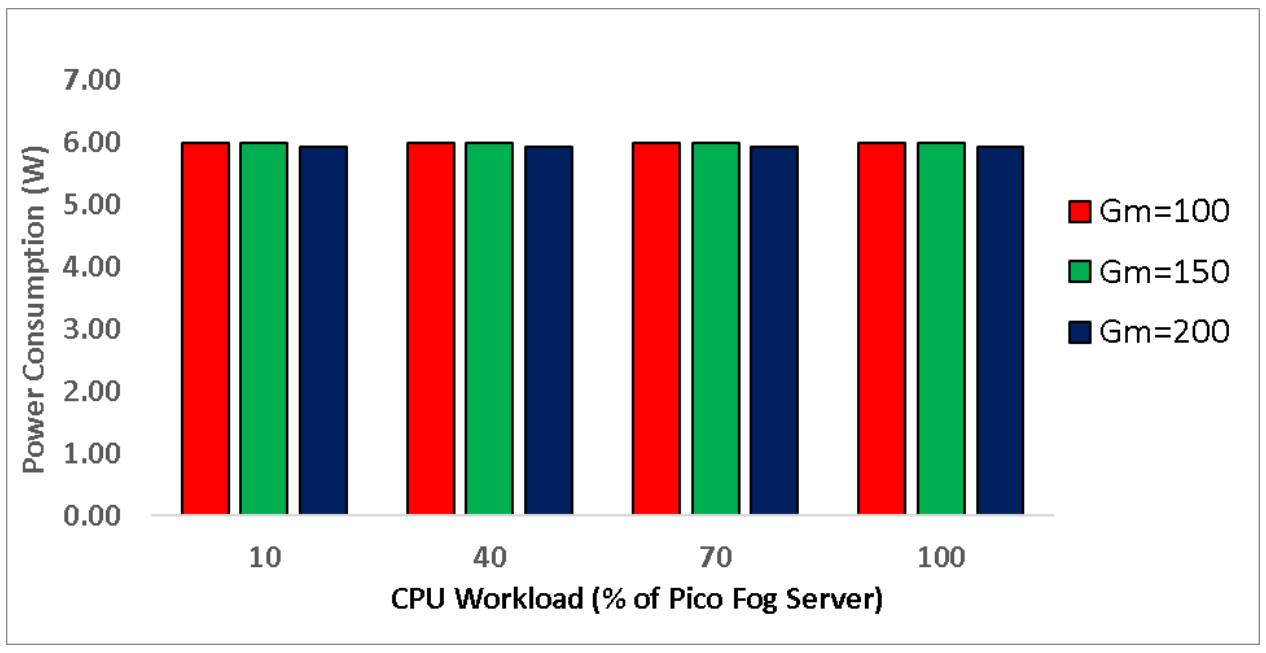

(b)

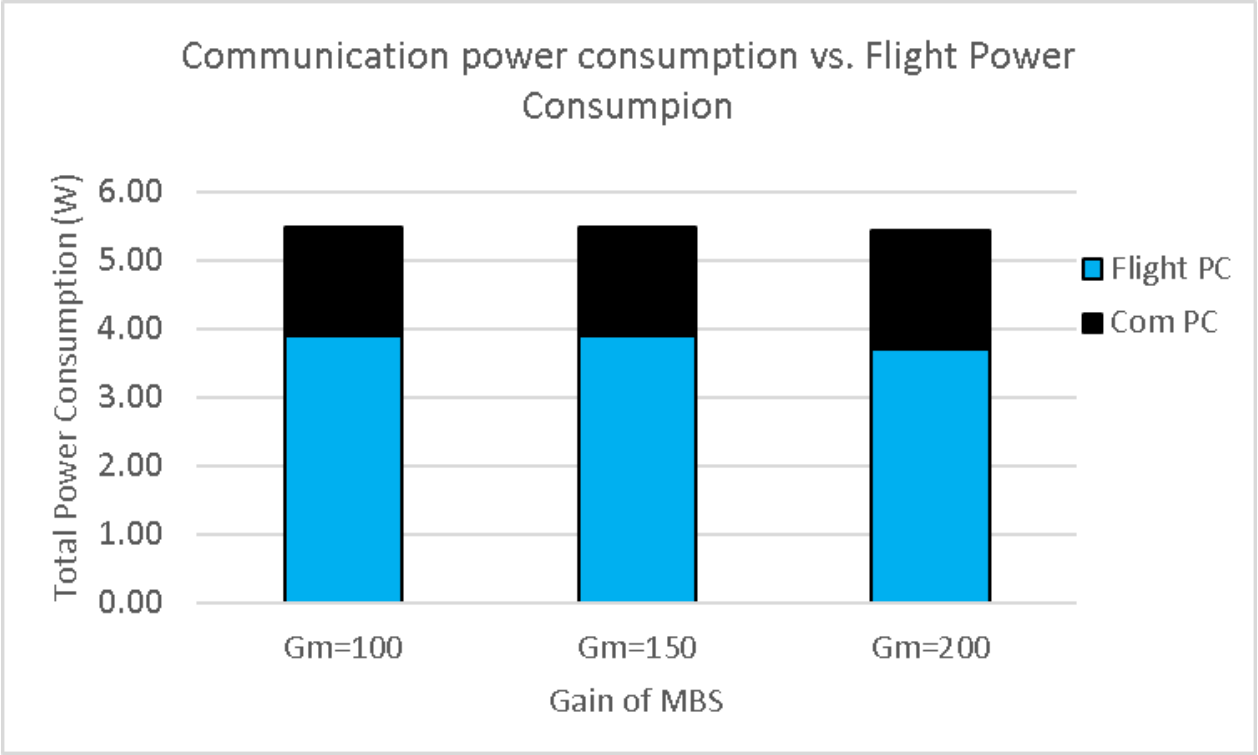

(c)

**FIGURE 4. Minimizing UAVTPC under a range of MBS gains and evaluating its impact on: (a) UAV's total flight distance, (b) UAV's total power consumption, (c) break down of UAV's total power consumption.**

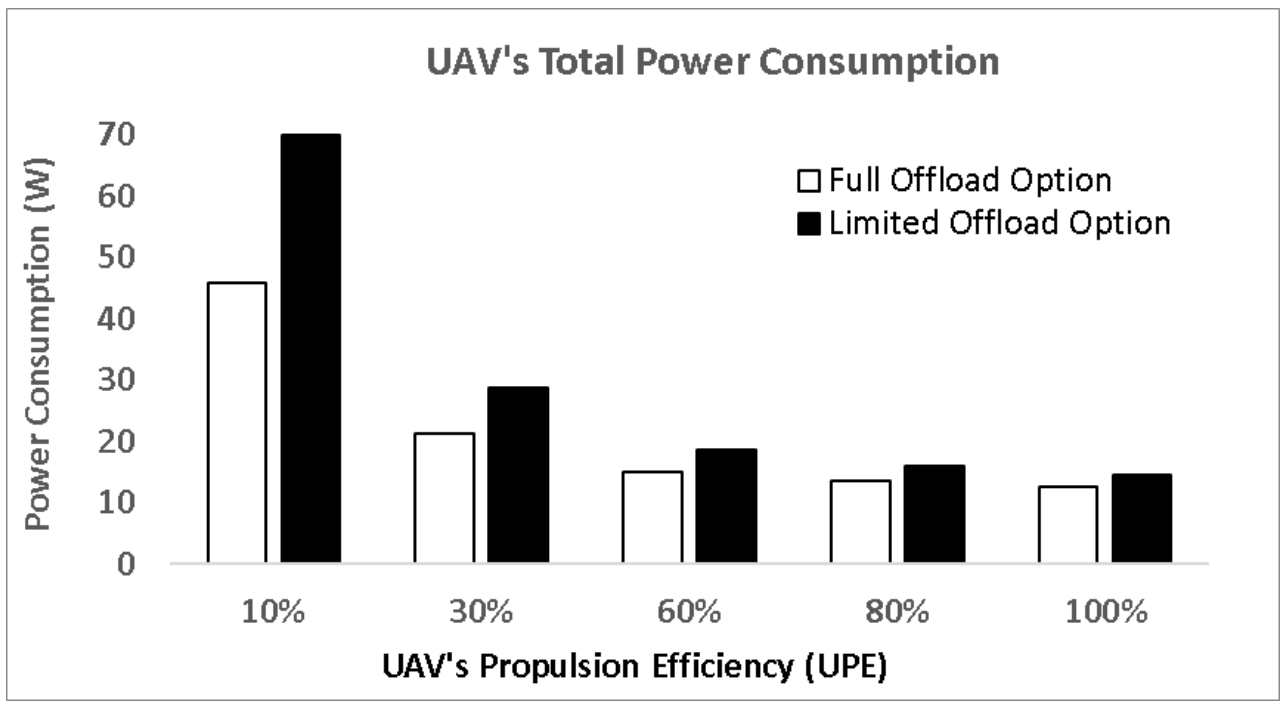

**FIGURE 5. UAV's total power consumption of the full offload and limited offload options against a range of UPE values.**

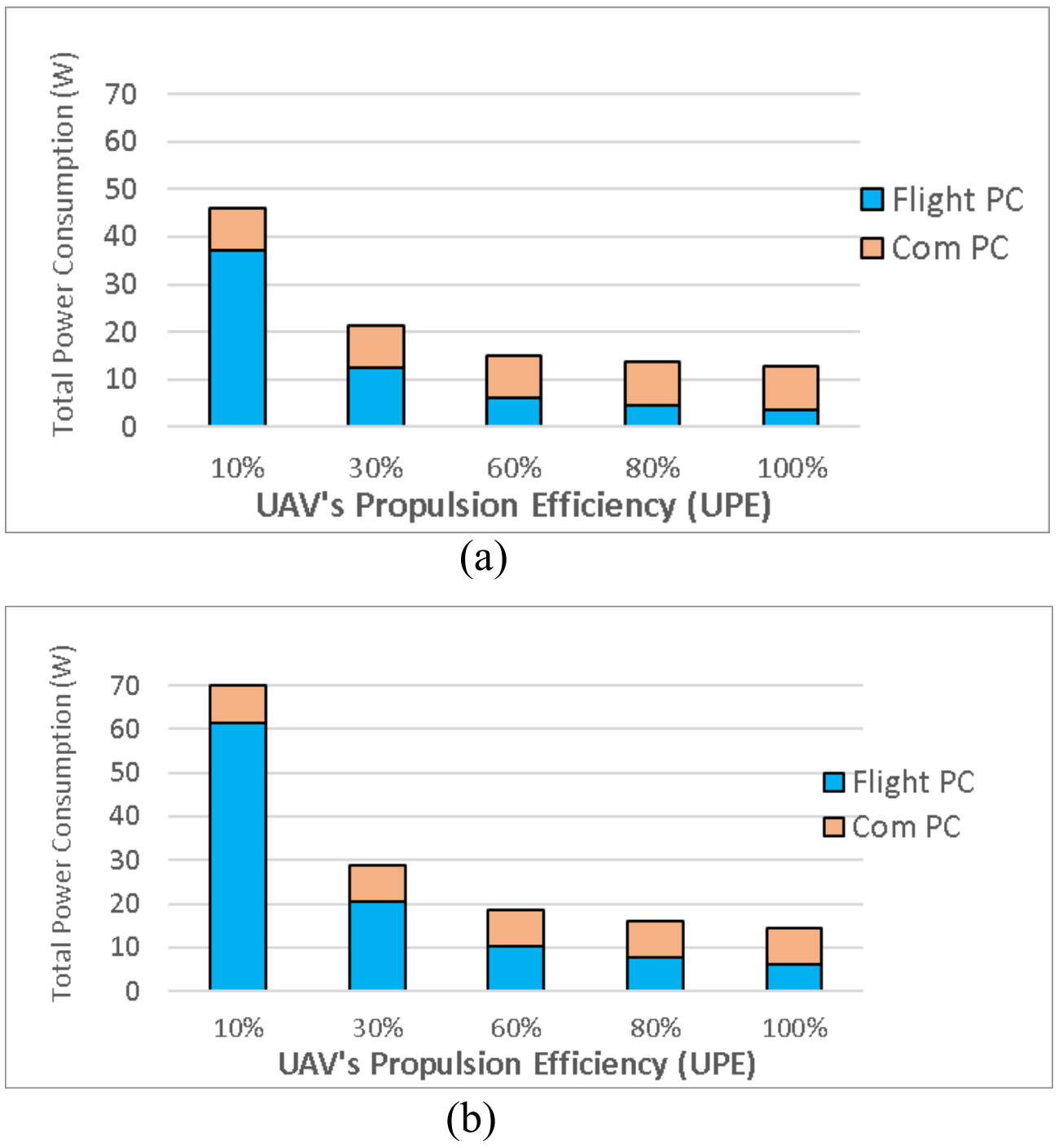

**FIGURE 6. Breakdown of UAVTPC into flight power consumption and communication power consumption under: (a) Full Offload Option, and (b) Limited Offload Option against a range of UPE values.**

## VI. CONCLUSIONS

This paper has investigated the joint optimization of UAV's power consumption and the computation offloading problem. A practical Cloud-Fog architecture is considered with a heterogenous wireless front-haul and a backhaul based on PON technologies is used to aggregate UAV traffic towards metro and cloud layers. We developed a multi-objective optimization model using MILP to minimize, individually, the power consumption of networking, power consumption of processing, power consumption of the UAV, and the total distance travelled by the UAV. We considered two test cases to examine 1) a scenario where Pico Fogs resources are in abundance and the UAV has an efficient propulsion, however the MBS's antenna gain is set to a typical value, and 2) a scenario that focused on the impact of propulsion efficiency on the offloading behaviour of the UAV. For future work, we intend to extend the current paper by designing heuristic algorithms that are better suited for real time implementations, considering a more sophisticated power consumption model of the UAV that better characterises the power consumption of different manoeuvres (e.g., ascending, descending, and hovering) in a 3D space and last but not least evaluating the service offloading problem for multi-UAV scenarios.

## ACKNOWLEDGMENTS

The authors extend their appreciation to the Deputyship for Research and Innovation, Ministry of Education in Saudi Arabia for funding this research work the project number (442/204). Also, the authors would like to extend their appreciation to Taibah University for its supervision support. For the purpose of open access, the authors have applied a Creative Commons Attribution (CC BY) license to any Author Accepted Manuscript version arising. All data are provided in full in the results section of this paper.

## REFERENCES

[1] R. Alyassi, M. Khonji, S. C.-K. Chau, K. Elbassioni, C.-M. Tseng, and A. Karapetyan, "Autonomous Recharging and Flight Mission Planning for Battery-operated Autonomous Drones," pp. 1–12, 2017, [Online]. Available: http://arxiv.org/abs/1703.10049.

[2] "Former Cisco CEO 500 Billion Connected Devices by 2025." Accessed: Apr. 16, 2022. [Online]. Available: https://www.businessinsider.com/former-cisco-ceo-500-billion-connected-devices-by-2025-2015-11?r=US&IR=T.

[3] A. Mohammed *et al.*, "DEEP REINFORCEMENT LEARNING FOR COMPUTATION OFFLOADING AND RESOURCE ALLOCATION IN BLOCKCHAIN-BASED MULTI-UAV-ENABLED MOBILE EDGE COMPUTING," pp. 295–299, 2020.

[4] O. K. Sahingoz, "Flyable Path Planning for a Multi-UAV System with Genetic Algorithms and Bezier Curves," pp. 41–48, 2013.

[5] M. Abrar, U. Ajmal, Z. M. Almohaimeed, X. Gui, R. Akram, and R. Masroor, "Energy Efficient UAV-Enabled Mobile Edge Computing for IoT Devices: A Review," *IEEE Access*, vol. 9, pp. 127779–127798, 2021, doi: 10.1109/ACCESS.2021.3112104.

[6] M. Samir, S. Sharafeddine, C. M. Assi, T. M. Nguyen, and A. Ghrayeb, "UAV Trajectory Planning for Data Collection from Time-Constrained IoT Devices," *IEEE Trans. Wirel. Commun.*, vol. 19, no. 1, pp. 34–46, 2020, doi: 10.1109/TWC.2019.2940447.

[7] M. Barcelo, A. Correa, J. Llorca, A. M. Tulino, J. L. Vicario, and A. Morell, "IoT-Cloud Service Optimization in Next Generation Smart Environments," vol. 34, no. 12, pp. 4077–4090, 2016.

[8] C. Mouradian, D. Naboulsi, S. Yangui, R. H. Glitho, M. J. Morrow, and P. A. Polakos, "A Comprehensive Survey on Fog Computing: State-of-the-Art and Research Challenges," *IEEE Commun. Surv. Tutorials*, vol. 20, no. 1, pp. 416–464, 2018, doi: 10.1109/COMST.2017.2771153.

[9] R. Mahmud, R. Kotagiri, and R. Buyya, "Fog Computing: A

Taxonomy, Survey and Future Directions," pp. 1–28, 2016, doi: 10.1007/978-981-10-5861-5_5.
[10] J. Li, J. Jin, D. Yuan, and H. Zhang, "Virtual Fog: A Virtualization Enabled Fog Computing Framework for Internet of Things," *IEEE Internet Things J.*, vol. 5, no. 1, pp. 121–131, 2018, doi: 10.1109/JIOT.2017.2774286.
[11] S. Sarkar, S. Chatterjee, and S. Misra, "Assessment of the Suitability of Fog Computing in the Context of Internet of Things," *IEEE Trans. Cloud Comput.*, vol. PP, no. 99, pp. 1–1, 2015, doi: 10.1109/TCC.2015.2485206.
[12] H. Guo, J. Liu, and H. Qin, "Collaborative Mobile Edge Computation Offloading for IoT over Fiber-Wireless Networks," *IEEE Netw.*, vol. 32, no. 1, pp. 66–71, Jan. 2018, doi: 10.1109/MNET.2018.1700139.
[13] B. A. Yosuf, M. Musa, T. Elgorashi, and J. Elmirghani, "Energy efficient distributed processing for IoT," *IEEE Access*. 2020.
[14] J. Almutairi, M. Aldossary, and H. A. Alharbi, "Delay-Optimal Task Offloading for UAV-Enabled Edge-Cloud Computing Systems," pp. 1–12, 2022.
[15] X. He, R. Jin, and H. Dai, "Multi-Hop Task Offloading with On-the-Fly Computation for Multi-UAV Remote Edge Computing," *IEEE Trans. Commun.*, vol. 70, no. 2, pp. 1332–1344, 2022, doi: 10.1109/TCOMM.2021.3129902.
[16] T. I. Applications, C. Zhan, H. Hu, Z. Liu, S. Member, and A. U. Uav, "Multi-UAV-Enabled Mobile-Edge Computing for," vol. 8, no. 20, pp. 15553–15567, 2021.
[17] R. Islambouli and S. Sharafeddine, "Optimized 3D Deployment of UAV-Mounted Cloudlets to Support Latency-Sensitive Services in IoT Networks," *IEEE Access*, vol. 7, pp. 172860–172870, 2019, doi: 10.1109/ACCESS.2019.2956150.
[18] C. Zhou *et al.*, "Delay-aware iot task scheduling in space-air-ground integrated network," *2019 IEEE Glob. Commun. Conf. GLOBECOM 2019 - Proc.*, 2019, doi: 10.1109/GLOBECOM38437.2019.9013393.
[19] S. Mao, S. He, and J. Wu, "Joint UAV Position Optimization and Resource Scheduling in Space-Air-Ground Integrated Networks With Mixed Cloud-Edge Computing," *IEEE Syst. J.*, vol. 15, no. 3, pp. 3992–4002, 2020, doi: 10.1109/JSYST.2020.3041706.
[20] C. Wang, W. Ding, Y. Luo, Y. Wang, J. Zhang, and J. Xiao, "Joint Trajectory Planning and Resource Allocation for UAV-assisted Information Collection," *2021 7th Int. Conf. Comput. Commun. ICCC 2021*, pp. 1068–1072, 2021, doi: 10.1109/ICCC54389.2021.9674487.
[21] J. Yao and N. Ansari, "Online Task Allocation and Flying Control in Fog-Aided Internet of Drones," *IEEE Trans. Veh. Technol.*, vol. 69, no. 5, pp. 5562–5569, 2020, doi: 10.1109/TVT.2020.2982172.
[22] X. He, R. Jin, and H. Dai, "Joint service placement and resource allocation for multi-UAV collaborative edge computing," *IEEE Wirel. Commun. Netw. Conf. WCNC*, vol. 2021-March, 2021, doi: 10.1109/WCNC49053.2021.9417565.
[23] Y. Yu, X. Bu, K. Yang, H. Yang, and Z. Han, "UAV-Aided Low Latency Mobile Edge Computing with mmWave Backhaul," *IEEE Int. Conf. Commun.*, vol. 2019-May, 2019, doi: 10.1109/ICC.2019.8761403.
[24] N. H. Motlagh, M. Bagaa, and T. Taleb, "Energy and Delay Aware Task Assignment Mechanism for UAV-Based IoT Platform," *IEEE Internet Things J.*, vol. 6, no. 4, pp. 6523–6536, 2019, doi: 10.1109/JIOT.2019.2907873.
[25] A. Paleti, "Performance Evaluation of Path Planning Techniques for Unmanned Aerial Vehicles A comparative analysis of A-star algorithm and Mixed Integer Linear Programming," 2016.
[26] A. Albert, F. S. Leira, and L. Imsland, "UAV path planning using MILP with experiments," *Model. Identif. Control*, vol. 38, no. 1, pp. 21–32, 2017, doi: 10.4173/mic.2017.1.3.
[27] S. Ahmed, A. Mohamed, K. Harras, M. Kholief, and S. Mesbah, "Energy efficient path planning techniques for UAV-based systems with space discretization," *IEEE Wirel. Commun. Netw. Conf. WCNC*, vol. 2016-Septe, no. Wcnc, 2016, doi: 10.1109/WCNC.2016.7565126.
[28] W. A. Kamal, D. W. Gu, and I. Postlethwaite, "MILP and its application in flight path planning," *IFAC Proc. Vol.*, vol. 38, no. 1, pp. 55–60, 2005, doi: 10.3182/20050703-6-cz-1902.02061.
[29] A. Ahmed, M. Naeem, and A. Al-Dweik, "Joint Optimization of Sensors Association and UAVs Placement in IoT Applications with Practical Network Constraints," *IEEE Access*, vol. 9, pp. 7674–7689, 2021, doi: 10.1109/ACCESS.2021.3049360.
[30] L. Huang, H. Qu, and L. Zuo, "Multi-Type UAVs Cooperative Task Allocation under Resource Constraints," *IEEE Access*, vol. 6, pp. 17841–17850, 2018, doi: 10.1109/ACCESS.2018.2818733.
[31] H. Xiao, Z. Hu, K. Yang, Y. Du, and D. Chen, "An Energy-Aware Joint Routing and Task Allocation Algorithm in MEC Systems Assisted by Multiple UAVs," *2020 Int. Wirel. Commun. Mob. Comput. IWCMC 2020*, pp. 1654–1659, 2020, doi: 10.1109/IWCMC48107.2020.9148519.
[32] Y. Zhu, S. Wang, X. Liu, H. Tong, and C. Yin, "Joint Task and Resource Allocation in SDN-based UAV-assisted Cellular Networks," *2020 IEEE/CIC Int. Conf. Commun. China, ICCC 2020*, no. Iccc, pp. 430–435, 2020, doi: 10.1109/ICCC49849.2020.9238969.
[33] X. Chen, T. Chen, Z. Zhao, H. Zhang, M. Bennis, and Y. Ji, "Resource Awareness in Unmanned Aerial Vehicle-Assisted Mobile-Edge Computing Systems," *IEEE Veh. Technol. Conf.*, vol. 2020-May, 2020, doi: 10.1109/VTC2020-Spring48590.2020.9128981.
[34] S. Zhang, J. Liu, Y. Zhu, and J. Zhang, "Joint Computation Offloading and Trajectory Design for Aerial Computing," *IEEE Wirel. Commun.*, vol. 28, no. 5, pp. 88–94, 2021, doi: 10.1109/MWC.011.2100073.
[35] X. Wei, C. Tang, J. Fan, and S. Subramaniam, "Joint optimization of energy consumption and delay in cloud-to-thing continuum," *IEEE Internet Things J.*, vol. 6, no. 2, pp. 2325–2337, 2019, doi: 10.1109/JIOT.2019.2906287.
[36] Y. Xu, T. Zhang, D. Yang, Y. Liu, and M. Tao, "Joint Resource and Trajectory Optimization for Security in UAV-Assisted MEC Systems," *IEEE Trans. Commun.*, vol. 69, no. 1, pp. 573–588, 2021, doi: 10.1109/TCOMM.2020.3025910.
[37] R. Duan, J. Wang, J. Du, C. Jiang, T. Bai, and Y. Ren, "Power-Delay Trade-off for Heterogenous Cloud Enabled Multi-UAV Systems," *IEEE Int. Conf. Commun.*, vol. 2019-May, 2019, doi: 10.1109/ICC.2019.8761377.
[38] K. Yao *et al.*, "Distributed Joint Optimization of Deployment, Computation Offloading and Resource Allocation in Coalition-based UAV Swarms," *12th Int. Conf. Wirel. Commun. Signal Process. WCSP 2020*, pp. 207–212, 2020, doi: 10.1109/WCSP49889.2020.9299672.
[39] B. Dai, J. Niu, T. Ren, Z. Hu, and M. Atiquzzaman, "Towards Energy-Efficient Scheduling of UAV and Base Station Hybrid Enabled Mobile Edge Computing," *IEEE Trans. Veh. Technol.*, vol. 71, no. 1, pp. 915–930, 2022, doi: 10.1109/TVT.2021.3129214.
[40] X. Ma, C. Yin, and X. Liu, "Machine Learning Based Joint Offloading and Trajectory Design in UAV Based MEC System for IoT Devices," *2020 IEEE 6th Int. Conf. Comput. Commun. ICCC 2020*, pp. 902–909, 2020, doi: 10.1109/ICCC51575.2020.9345069.
[41] X. Huang, X. Yang, Q. Chen, and J. Zhang, "Task Offloading Optimization for UAV-Assisted Fog-Enabled Internet of Things Networks," *IEEE Internet Things J.*, vol. 9, no. 2, pp. 1082–1094, 2022, doi: 10.1109/JIOT.2021.3078904.
[42] O. M. Rosabal, O. A. Lopez, D. E. Perez, M. Shehab, H. Hilleshein, and H. Alves, "Minimization of the Worst-Case Average Energy Consumption in UAV-Assisted IoT Networks," *IEEE Internet Things J.*, pp. 1–1, 2022, doi: 10.1109/jiot.2022.3150419.
[43] B. J. Baliga, R. W. A. Ayre, K. Hinton, R. S. Tucker, and F. Ieee, "Green Cloud Computing : Balancing Energy in Processing , Storage , and Transport," 2011.
[44] F. Jalali, S. Khodadustan, C. Gray, K. Hinton, and F. Suits, "Greening IoT with Fog: A Survey," in *Proceedings - 2017 IEEE 1st International Conference on Edge Computing, EDGE 2017*, Sep. 2017, pp. 25–31, doi: 10.1109/IEEE.EDGE.2017.13.
[45] F. Jalali, K. Hinton, R. Ayre, T. Alpcan, and R. S. Tucker, "Fog computing may help to save energy in cloud computing," *IEEE J.*

*Sel. Areas Commun.*, vol. 34, no. 5, pp. 1728–1739, 2016, doi: 10.1109/JSAC.2016.2545559.

[46] G. S. G. Shen and R. S. Tucker, "Energy-Minimized Design for IP Over WDM Networks," *IEEE/OSA J. Opt. Commun. Netw.*, vol. 1, no. 1, pp. 176–186, 2009, doi: 10.1364/JOCN.1.000176.

[47] W. Yu *et al.*, "A Survey on the Edge Computing for the Internet of Things," *IEEE Access*, vol. 6, pp. 6900–6919, 2017, doi: 10.1109/ACCESS.2017.2778504.

[48] Z. T. Al-Azez, A. Q. Lawey, T. E. H. El-Gorashi, and J. M. H. Elmirghani, "Energy Efficient IoT Virtualization Framework With Peer to Peer Networking and Processing," *IEEE Access*, vol. 7, pp. 50697–50709, 2019, doi: 10.1109/ACCESS.2019.2911117.

[49] F. A. Salaht, F. Desprez, and A. Lebre, "An Overview of Service Placement Problem in Fog and Edge Computing," *ACM Comput. Surv.*, vol. 53, no. 3, 2020, doi: 10.1145/3391196.

[50] F. Jalali, K. Hinton, R. Ayre, T. Alpcan, and R. S. Tucker, "Fog computing may help to save energy in cloud computing," *IEEE J. Sel. Areas Commun.*, vol. 34, no. 5, pp. 1728–1739, 2016, doi: 10.1109/JSAC.2016.2545559.

[51] C. Gray, R. Ayre, K. Hinton, and R. S. Tucker, "Power consumption of IoT access network technologies," *2015 IEEE Int. Conf. Commun. Work.*, pp. 2818–2823, 2015, doi: 10.1109/ICCW.2015.7247606.

[52] A. N. Al-Quzweeni, A. Q. Lawey, T. E. H. Elgorashi, and J. M. H. Elmirghani, "Optimized Energy Aware 5G Network Function Virtualization," *IEEE Access*, vol. 7, pp. 44939–44958, 2019, doi: 10.1109/ACCESS.2019.2907798.

[53] I. Corporation and B. Sig, "Inspiron 2350 AIO Specifications," 2013.

[54] "7-Zip Compression Benchmark - OpenBenchmarking.org." https://openbenchmarking.org/test/pts/compress-7zip (accessed Apr. 21, 2022).

[55] A. Shehabi *et al.*, "United States Data Center Energy Usage Report," *Lawrence Berkeley Natl. Lab. Berkeley, CA, Tech. Rep.*, no. June, pp. 1–66, 2016, doi: LBNL-1005775.

[56] P. Sarigiannidis, T. Lagkas, S. Bibi, A. Ampatzoglou, and P. Bellavista, "Hybrid 5G optical-wireless SDN-based networks, challenges and open issues," *IET Networks*, vol. 6, no. 6, pp. 141–148, 2017, doi: 10.1049/iet-net.2017.0069.

[57] N. Obaid and A. Czylwik, "The impact of deploying pico base stations on capacity and energy efficiency of heterogeneous cellular networks," *IEEE Int. Symp. Pers. Indoor Mob. Radio Commun. PIMRC*, pp. 1904–1908, 2013, doi: 10.1109/PIMRC.2013.6666454.

[58] D. Meisner, B. T. Gold, and T. F. Wenisch, *PowerNap: Eliminating Server Idle Power.* .

[59] "Mavic Pro - Product Information - DJI." https://www.dji.com/uk/mavic/info (accessed Apr. 23, 2022).

[60] H. V. Abeywickrama, B. A. Jayawickrama, Y. He, and E. Dutkiewicz, "Empirical Power Consumption Model for UAVs," *IEEE Veh. Technol. Conf.*, vol. 2018-Augus, pp. 5–9, 2018, doi: 10.1109/VTCFall.2018.8690666.

[61] N. Gao *et al.*, "Energy model for UAV communications: Experimental validation and model generalization," *China Commun.*, vol. 18, no. 7, pp. 253–264, 2021, doi: 10.23919/JCC.2021.07.020.

[62] J. Huang, Y. Meng, X. Gong, Y. Liu, and Q. Duan, "A Novel Deployment Scheme for Green Internet of Things," *IEEE Internet Things J.*, vol. PP, no. 99, pp. 1–1, 2014, doi: 10.1109/JIOT.2014.2301819.

[63] A. M. Al-Salim, A. Q. Lawey, T. E. H. El-Gorashi, and J. M. H. Elmirghani, "Energy Efficient Big Data Networks: Impact of Volume and Variety," *IEEE Trans. Netw. Serv. Manag.*, vol. 15, no. 1, pp. 1–1, 2017, doi: 10.1109/TNSM.2017.2787624.

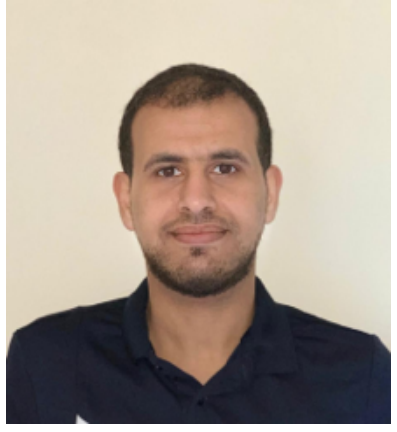

**Hatem A. Alharbi** received the B.Sc. degree in Computer Engineering (Hons.) from Umm Alqura University, Makkah, Saudi Arabia in 2012, the M.Sc. degree in Digital communication networks (with distinction) in 2015, and the PhD degree in communication networks in 2020, from the University of Leeds, Leeds, UK. Dr. Alharbi is currently an assistant professor in the Computer Engineering Department in the School of Computer Science and Engineering, University of Taibah, Saudi Arabia. His research interests are in energy efficient fog and cloud networks and network economics.

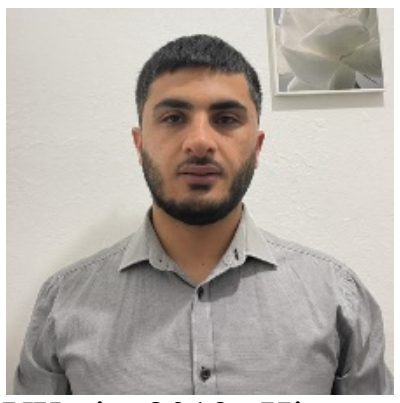

**Barzan A. Yosuf** received the BEng (Hons) degree in computer systems engineering, in 2012 and the MSc degree in embedded systems engineering in 2013, from the University of Huddersfield, Huddersfield, UK, and the Ph.D. degree with the School of Electronic and Electrical Engineering, in energy efficient distributed processing for IoT from the University of Leeds, UK, in 2019. His current research interests include IoT, fog and cloud computing energy optimization. He is currently a member of the IEEE Energy Efficient ICT working group, which focuses on developing improved energy efficient standards in areas such as big data processing over core networks, network virtualization and VM placement in the presence of access fog and core clouds.

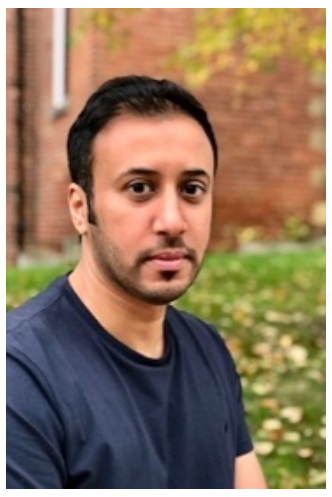

**Mohammad Aldossary** received his B.Sc. degree in Computer (Hons.) from King Saud University, Riyadh, Saudi Arabia in 2009 and the M.Sc. degree in Computer Science from Southern Polytechnic State University, Georgia, USA, in 2013. He was awarded a Ph.D. degree in Computer Science from the University of Leeds, UK in 2019. His main research areas focus on Distributed Systems, including Cloud Computing, Fog Computing, Edge Computing, Internet of Things, System Architectures, Resource Management, and Energy Efficiency. Dr Aldossary is currently an Assistant Professor in the Computer Science Department at Prince Sattam bin Abdulaziz University, Saudi Arabia.

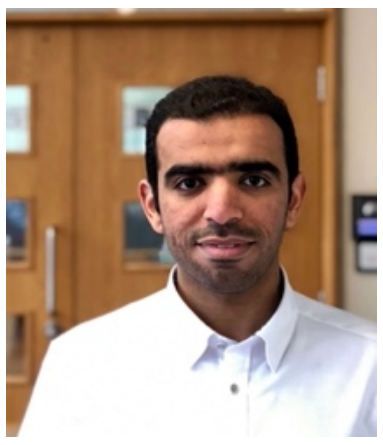

**jaber Almutairi** received his BSc degree in Computer Science from Taibah University, Medina, Saudi Arabia in 2012 and his MSc degree in Advanced Computer Science (Cloud Computing) and PhD from University of Leeds, Leeds, UK in 2016 and 2020. His research interests include resource management and simulation in the edge computing environment and applications of the Internet of Things. He is currently Assistant Professor with the Department of Computer Science, College of Computer Science and Engineering, Taibah University, Madinah, Saudi Arabia.

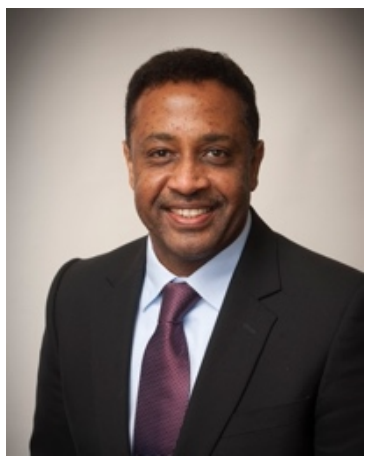

**Professor Jaafar Elmirghani** is Fellow of IEEE, Fellow of the IET, Fellow of the Institute of Physics and is the Director of the Institute of Communication and Power Networks and Professor of Communication Networks and Systems within the School of Electronic and Electrical Engineering, University of Leeds, UK. He joined Leeds in 2007 having been full professor and chair in Optical Communications at the University of Wales Swansea 2000-2007.

He received the BSc in Electrical Engineering, First Class Honours from the University of Khartoum in 1989 and was awarded all 4 prizes in the department for academic distinction. He received the PhD in the synchronization of optical systems and optical receiver design from the University of Huddersfield UK in 1994 and the DSc in Communication Systems and Networks from University of Leeds, UK, in 2014. He co-authored Photonic Switching Technology: Systems and Networks, (Wiley) and has published over 550 papers.

He was Chairman of the IEEE UK and RI Communications Chapter and was Chairman of IEEE Comsoc Transmission Access and Optical Systems Committee and Chairman of IEEE Comsoc Signal Processing and Communication Electronics (SPCE) Committee. He was a member of IEEE ComSoc Technical Activities Council' (TAC), is an editor of IEEE Communications Magazine and is and has been on the technical program committee of 43 IEEE ICC/GLOBECOM conferences between 1995 and 2022 including 21 times as Symposium Chair. He was founding Chair of the Advanced Signal Processing for Communication Symposium which started at IEEE GLOBECOM'99 and has continued since at every ICC and GLOBECOM. Prof. Elmirghani was also founding Chair of the first IEEE ICC/GLOBECOM optical symposium at GLOBECOM'00, the Future Photonic Network Technologies, Architectures and Protocols Symposium. He chaired this Symposium, which continues to date. He was the founding chair of the first Green Track at ICC/GLOBECOM at GLOBECOM 2011, and is Chair of the IEEE Sustainable ICT Initiative, a pan IEEE Societies Initiative responsible for Green ICT activities across IEEE, 2012-present. He has given over 90 invited and keynote talks over the past 15 years.

He received the IEEE Communications Society 2005 Hal Sobol award for exemplary service to meetings and conferences, the IEEE Communications Society 2005 Chapter Achievement award, the University of Wales Swansea inaugural 'Outstanding Research Achievement Award', 2006, the IEEE Communications Society Signal Processing and Communication Electronics outstanding service award, 2009, a best paper award at IEEE ICC'2013, the IEEE Comsoc Transmission Access and Optical Systems outstanding Service award 2015 in recognition of "Leadership and Contributions to the Area of Green Communications", the GreenTouch 1000x award in 2015 for "pioneering research contributions to the field of energy efficiency in telecommunications", the IET 2016 Premium Award for best paper in IET Optoelectronics, shared the 2016 Edison Award in the collective disruption category with a team of 6 from GreenTouch for their joint work on the GreenMeter, the IEEE Communications Society Transmission, Access and Optical Systems technical committee 2020 Outstanding Technical Achievement Award for outstanding contributions to the "energy efficiency of optical communication systems and networks". He was named among the top 2% of scientists in the world by citations in 2019 in Elsevier Scopus, Stanford University database which includes the top 2% of scientists in 22 scientific disciplines and 176 sub-domains. Named also in the following year (2021) edition. He was elected Fellow of IEEE for "Contributions to Energy-Efficient Communications," 2021.

He is currently an Area Editor of IEEE Journal on Selected Areas in Communications series on Machine Learning for Communications, an editor of IEEE Journal of Lightwave Technology, IET Optoelectronics and Journal of Optical Communications, and was editor of IEEE Communications Surveys and Tutorials and IEEE Journal on Selected Areas in Communications series on Green Communications and Networking. He was Co-Chair of the GreenTouch Wired, Core and Access Networks Working Group, an adviser to the Commonwealth Scholarship Commission, member of the Royal Society International Joint Projects Panel and member of the Engineering and Physical Sciences Research Council (EPSRC) College.

He has been awarded in excess of £30 million in grants to date from EPSRC, the EU and industry and has held prestigious fellowships funded by the Royal Society and by BT. He was an IEEE Comsoc Distinguished Lecturer 2013-2016. He was PI of the £6m EPSRC Intelligent Energy Aware Networks (INTERNET) Programme Grant, 2010-2016 and is currently PI of the EPSRC £6.6m Terabit Bidirectional Multi-user Optical Wireless System (TOWS) for 6G LiFi, 2019-2024. He leads a number of research projects and has research interests in energy efficiency, communication networks, wireless and optical communication systems.